\documentclass[showpacs,amsmath,amssymb,aps]{revtex4} 
\input{epsf}
\usepackage{color,graphicx}
\usepackage{natbib}
\def\beq{\begin{equation}}
\def\eeq{\end{equation}}
\def\bea{\begin{eqnarray}}
\def\eea{\end{eqnarray}}
\def\n{{\rm n}}
\def\e{{\rm e}}
\def\p{{\rm p}}

\def\A{{{\cal A}^{\x \y}}}
\def\B{{{\cal B}^\x}}
\def\x{{\rm x}}
\def\y{{\rm y}}
\def\X{{\cal X}^{\x \y}}
\def\a{{\rm A}}
\def\b{{\rm B}}
\def\d{{\rm D}}
\def\wx{{W_\x}}
\def\vx{{V_\x}}
\def\xx{{X_\x}}
\def\Ha{{H_0}}
\def\H1{{H_1}}

\def\ms{{m_*}}
\def\tH1{\tilde{H_1}}
\def\tK{\tilde{K}}
\def\tWp{\tilde{W}}
\def\tXp{\tilde{X}}
\def\tp{\tilde{n}}
\def\tchi{\tilde{\mu}}

\begin{document}
\title{Oscillations of General Relativistic Multi-fluid/Multi-layer Compact 
Stars}

\author{Lap-Ming~Lin}
\email{lmlin@phy.cuhk.edu.hk}
\affiliation{Department of Physics and Institute of Theoretical Physics, 
The Chinese University of Hong Kong, Hong Kong, China}

\author{N.~Andersson}
\email{na@maths.soton.ac.uk}
\affiliation{Department of Mathematics, University of Southampton, 
Southampton SO17 1BJ, UK}

\author{G.~L.~Comer}
\email{comergl@slu.edu}
\affiliation{Department of Physics, Saint Louis University, St.~Louis, 
MO, 63156-0907, USA}

\date{\today}

\begin{abstract}
We develop the formalism for determining the quasinormal modes of general 
relativistic multi-fluid compact stars in such a way that the impact of 
superfluid gap data can be assessed. \ Our results represent the first  
attempt to study true multi-layer dynamics, an important step towards 
considering realistic superfluid/superconducting compact stars. \ We combine 
a relativistic model for entrainment with model equations of state that 
explicity incorporate the symmetry energy. \ Our analysis emphasises the many 
different parameters that are required for this kind of modelling, and the 
fact that standard tabulated equations of state are grossly incomplete in 
this respect. \ To make progress, future equations of state need to provide
the energy density as a function of the various nucleon number densities, the 
temperature (i.e.~entropy), and the entrainment among the various components.
\end{abstract}


\maketitle

\section{Introduction}
The fluid approximation \cite{andersson07:_livrev} is a necessity for 
modeling systems containing so many elementary constituents that (on a 
macroscopic scale) they form a continuum, and can redistribute energy and 
momentum among themselves. \ The model is based on the notion of fluid 
element, which is small enough to be infinitesimal with respect to the system 
${en~masse}$, but large enough to contain, say, an Avogadro's number worth of 
particles. \ In a superfluid system one must consider different, dynamically 
decoupled, yet co-existing fluids. \ Each individual fluid has its own 
collection of fluid elements and each spacetime point in the system will have 
as many fluid element worldlines passing through it as there are independent 
fluids. \ The extent to which different fluids are coupled depends largely on 
the available dissipation mechanisms, eg.~friction due to interparticle 
scattering. \ In a multi-fluid system this is a complex problem 
\cite{andersson05:_flux_con}. \ Generally, the system dissipation depends on 
how energy and momentum are exchanged as the fluid elements expand and 
contract, how they slide across each other, how they rotate about each other, 
and how they flow through each other. 

Neutron stars are believed to be prime examples of general relativistic, 
multi-fluid objects. \ To date, on the order of a couple of thousand pulsars 
have been observed \cite{lorimer05:_lrr}. Yet, an extrapolation of the local 
population data for our galaxy suggests the existence of about 
$1.6 \times 10^5$ normal pulsars, around $4 \times 10^4$ millisecond pulsars, 
and about $2 \times 10^8$ neutron stars that are no longer active pulsars. \ 
Most of these should be extremely cold in the sense that their temperature is 
several orders of magnitude less than the Fermi temperatures of the 
independent (massive) particle species, i.e.~$10^{12}~{\rm K}$. \ In fact,  
neutron stars cool to temperatures below $10^9~{\rm K}$ relatively soon after 
birth. \ This is the expected transition temperature 
\cite{lombardo99:_conf,lombardo01:_lnp,andersson04:_viscosity_coeff} 
for neutrons and protons to become superfluid and superconducting, 
respectively. \ We therefore anticipate that most neutron stars in our galaxy 
have at least two superfluids/superconductors in their cores. \ In fact, 
neutron superfluidity is a key ingredient in most models of large pulsar 
glitches \cite{radha69:_glitches,lyne93:_conf}, with (catastrophic) transfer 
of angular momentum via vortices from a superfluid component to the crust 
leading to the observed spin-up.

Ever since neutron star superfluidity was first suggested \cite{migdal59:_np}, 
we have seen a concerted effort aimed at developing our understanding of the 
various phases of matter 
\cite{lombardo99:_conf,lombardo01:_lnp,clark92:_conf,glendenning97:_compact_stars,Walecka:1995mi}. 
\ This has led to a picture where a typical neutron star has a number of 
distinct ``layers''. \ From the outer to the inner crust, protons are locked 
inside increasingly neutron-rich nuclei which are embedded in a degenerate 
normal fluid of electrons. \ At the base of the crust, the nuclei are 
believed to assume exotic ``pasta'' shapes \cite{pethick95:_rev}. \ Moreover, 
in the crust region the long-range, attractive component of the nuclear force 
should lead to Cooper pairing of neutrons in ${}^1{\rm S}_0$ states. \ The 
crust nuclei will thus be embedded in and penetrated by superfluid neutrons. 
\ Short-range repulsion in the nuclear force and the spin-orbit interaction 
allow neutrons to pair as ${}^3{\rm P}_2$ states in the more dense regions of 
the outer core \cite{hoffberg70:_prl}. \ There are no nuclei in this region 
and the protons that remain are dilute enough to feel only the long-range 
attractive part of the nuclear force. Hence, they are expected to pair in 
${}^1{\rm S}_0$ states. \ The core neutrons and protons are embedded in a 
highly degenerate normal fluid of electrons. \ At high enough density it 
becomes energetically favorable for more massive particles to form (eg.~muons 
in lieu of electrons). \ At the most extreme densities, quarks may become 
deconfined, possibly opening up a number of channels for attractive 
interactions and hence many Cooper pairing possibilities (e.g.~so-called CFL 
matter \cite{alford04:_review,alford00:_cfl}). \ Each individual Cooper 
pairing will lead to individual condensates, and potentially many 
inter-penetrating fluids. \ The actual number of dynamically distinct fluids 
depends very much on the details of dissipation 
\cite{andersson05:_flux_con,Gusakov07:_vis}. \ The scattering time-scale must 
typically be much greater than the characteristic dynamical time-scales in 
order for fluid components to decouple.

A moderately realistic neutron star model must account for the presence
of different regions in the star. \ Describing a neutron star as a multi-layer 
system, one should at the very least account for the presence of the 
(multi-fluid) superfluid/superconducting regions and the elastic crust 
component. \ Even though there has been progress in this area, we are still 
not at this level of sophistication. \ The current state-of-the-art is 
represented by \cite{andersson02:_oscil_GR_superfl_NS}, where quasinormal 
modes for a  core-envelope model are calculated, and a recent study of axial 
crust oscillations in the relativistic Cowling approximation 
\cite{Samuelsson07:_axial_crust}. \ The aim of the present work is to improve 
on the first aspect of the modeling. \ We will include, for the first time, 
the physics of the superfluid phase transition in the construction of the 
spherically symmetric and static background model. \ The motivation for this 
is that the different regions of superfluidity may not overlap 
\cite{andersson04:_viscosity_coeff}, and the fluid dynamics will change 
depending on whether a given layer is normal or superfluid. \ This represents 
an important improvement on previous models, since it allows us to quantify 
how changes in the superfluid energy gap affect the quasinormal-mode 
spectrum. \ We leave inclusion of the crust elasticity for future studies. \ 
A fully relativistic formalism for a crust penetrated by a superfluid has 
been developed, see eg.~\cite{carter06:_crust}, but it is yet to be applied 
to neutron star dynamics. 

In a mixture of the superfluids ${\rm He}^3$ and ${\rm He}^4$, it is known 
that a momentum induced in one of the constituents will cause some of the 
mass of the other to be carried along, or entrained 
\cite{andreev75:_three_velocity_hydro,vardanyan81:_entrain}. \ Another 
example is the entropy (which can be considered as a massless fluid). \ In 
fact, in superfluid ${\rm He}^4$ the so-called ``normal'' fluid density is 
directly proportional to the entrainment between the atoms and the entropy. 
\ In a neutron star, the strong interaction leads to entrainment between 
neutrons and protons. \ Entrainment is a multi-fluid effect, that has no 
counterpart in a single-fluid system. \ When the neutrons start to flow they 
will, through entrainment, induce a momentum in the protons and subsequently 
the electrons \cite{alpar84:_rapid_postglitch}. \ In principle, there could 
be entrainment between each and every fluid of a multi-fluid system. \ In 
this analysis we will only consider the entrainment between neutrons and 
protons, for which we use the fully relativistic mean field model developed 
by Comer and Joynt \cite{comer03:_rel_ent}.

The outline of this paper is as follows: In Sec.~\ref{review} we give details 
on the two-fluid formalism, and the set of equations used to model the 
quasinormal modes. \ Sec.~\ref{anexp} discusses an expansion for the local 
matter content that is adapted to include entrainment at the appropriate 
order. \ Sec.~\ref{masfcn} gives the specifics on the local matter content, 
i.e.~the equation of state, the gap data, and the relativistic 
entrainment. \ The following Sec.~\ref{results} provides the results of our 
analysis of modes for two equations of state, with or without entrainment, 
and different ``temperatures'' of the star. \ Sec.~\ref{neweos} gives some 
concluding remarks and discusses to what extent the extant literature and 
computational infrastructure for compact object equations of state are 
adequate for supporting our kind of analysis. \ Finally, the Appendix 
describes the technique used to obtain the results in Sec.~\ref{results}. \ 
We use ``MTW'' conventions throughout.

\section{General Relativistic Two-fluid Formalism} \label{review} 

\subsection{The full formalism}

We will use the formalism developed by Carter, Langlois, and their various 
collaborators 
\cite{carter89:_covar_theor_conduc,comer93:_hamil_multi_con,comer94:_hamil_sf,carter95:_kalb_ramond,langlois98:_differ_rotat_superfl_ns,comer99:_quasimodes_sf,prix00:_cov_vortex,andersson01:_slowl_rotat_GR_superfl_NS,comer02:_zero_freq_subspace,prix04:_multi_fluid} (see Andersson and Comer 
\cite{andersson07:_livrev} for a recent review). \ The fundamental fluid 
variables consist of the conserved nucleon density four-currents, to be 
denoted $n^\mu_\x$ where $\x = \{\n,\p\}$ is a so-called constituent index 
(which is not summed over when repeated). \ From the currents can be formed 
three scalars: $n^2_\n = - g_{\mu \nu} n^\mu_\n n^\nu_\n$, $n^2_\p = - 
g_{\mu \nu} n^\mu_\p n^\nu_\p$, and $n^2_{\n \p} = - g_{\mu \nu} n^\mu_\n 
n^\nu_\p$. \ Given a master function $- \Lambda(n^2_\n,n^2_\p,n^2_{\n \p})$ 
(the two-fluid analog of the equation of state), the stress-energy tensor is  
\beq
    T^\mu{}_\nu = \Psi \delta^\mu{}_\nu + n^\mu_\n \mu^\n_\nu + n^\mu_\p 
                  \mu^\p_\nu \ ,
\eeq
where 
\beq
    \Psi = \Lambda - n^\rho_\n \mu^\n_\rho - n^\rho_\p \mu^\p_\rho 
           \label{press}
\eeq
is the generalized pressure and
\beq
    \mu^\x_\nu = g_{\mu \nu} \left(\B n^\nu_\x + \A n^\nu_\y\right) \ , 
\eeq
is the chemical potential covector. \ It is also the momentum canonically 
conjugate to the current $n^\mu_\x$. \ Formally, the $\A$ and $\B$ 
coefficients are obtained from the master function via the partial 
derivatives
\beq
    \A = {\cal A}^{\y \x} = - \frac{\partial \Lambda}{\partial n^2_{\x \y}} 
         \quad , \quad 
    \B = - 2 \frac{\partial \Lambda}{\partial n^2_\x} \ . \quad 
\eeq
The fact that the momentum $\mu^\x_\mu$ is not simply proportional to the
corresponding number density current $n^\mu_\x$ is a result of entrainment. \ 
It vanishes if the $\A$ coefficient is zero.

Finally, the equations for the neutrons and protons each consist of a 
conservation equation  
\beq
    \nabla_{\mu} n^{\mu}_\x = 0 \ , \label{conteq}
\eeq
and an Euler equation 
\beq
    n^{\mu}_\x \omega^\x_{\mu \nu} = 0 \ , \label{euler}
\eeq
where the vorticity two-form is defined by
\beq
    \omega^\x_{\mu \nu} = 2 \nabla_{[\mu} \mu^\x_{\nu]} \ .
\eeq
The square brackets indicate antisymmetrization of the enclosed indices. \ 
Comer \cite{comer02:_zero_freq_subspace} and Prix et al 
\cite{prix02:_slow_rot_ns_entrain} discuss in some detail why the assumption 
of separate conservation laws for the two fluids should be reasonable for slow 
rotation and quasinormal-mode calculations (excluding the thin transition 
layers that separate single fluid and multi-fluid layers). \ Note that the 
above way of writing the Euler equations makes manifest its geometric meaning 
as an integrability condition for the vorticity, a point that has been much 
emphasized by Carter \cite{carter89:_covar_theor_conduc} (see also 
\cite{andersson07:_livrev}).

\subsection{Equilibrium models}

In order to determine the background fluid configuration we need to evaluate 
the associated metric. \ We take our equilibrium configurations to be static 
and spherically symmetric. \ The metric can thus be written in the 
Schwarzschild form
\beq
    ds^2 = - e^{\nu} dt^2 + e^{\lambda}  dr^2 + r^2 \left(d\theta^2 + 
           {\rm sin}^2\theta d\phi^2\right) \ . \label{bgmet}
\eeq
The required metric coefficients are determined from two components of the 
Einstein equations, which can be written  
\beq
    \lambda^{\prime} = \frac{1 - e^{\lambda}}{r} - 8 \pi r 
                         e^{\lambda} \Lambda \quad , \quad
    \nu^{\prime} = - \frac{1 - e^{\lambda}}{r} + 8 \pi r 
                         e^{\lambda} \Psi \ . \label{bckgrnd}
\eeq
A prime represents a radial derivative, and it is to be understood that 
$\Lambda = \Lambda(n^2_\n,n^2_\p)$ and $\Psi = \Psi(n^2_\n,n^2_\p)$ in the 
two-fluid layers and $\Lambda = \Lambda(n^2)$ and $\Psi = \Psi(n^2)$, where 
$n=n_\n+n_\p$ is the total baryon number density, in the single-fluid layers.  

The equation that determines the radial profile of $n_\x(r)$ in the 
superfluid layer is \cite{comer99:_quasimodes_sf}  
\beq
    \b^\x{}^0_0 n_\x^{\prime} + \a^{\x \y}{}^0_0 n_\y^{\prime} + 
           \frac{1}{2} \mu^\x \nu^{\prime} = 0 \ , \label{nprof}
\eeq
where
\begin{eqnarray}
    \a^{\x \y}{}_0^0 &=& {\cal A}^{\x \y} + 2 \frac{\partial 
          {\cal B}^\x}{\partial n_\y^2} n_\x n_\y + 2 
          \frac{\partial {\cal A}^{\x \y}}{\partial n_\x^2} n_\x^2 + 2 
          \frac{\partial {\cal A}^{\x \y}}{\partial n_\y^2} n_\y^2 + 
          \frac{\partial {\cal A}^{\x \y}}{\partial n^2_{\x \y}} n_\y n_\x 
          \ , \\
          && \cr
    \b^\x{}_0^0 &=& \B + 2 \frac{\partial \B}{\partial n_\x^2} 
          n_\x^2 + 4 \frac{\partial \A}{\partial n_\x^2} n_\x n_\y + 
          \frac{\partial \A}{\partial n_{\x \y}^2} n_\y^2 \ , \label{coef2}
\end{eqnarray}
and $\mu^\x = \mu^\x_0$ is the background chemical potential. \ In evaluating 
these coefficients, one sets $n^2_{\n \p} = n_\n n_\p$ after the partial 
derivatives are taken. 

In the single-fluid layers the only matter equation concerns the nucleon 
radial profile $n(r)$:
\beq
    \b^0_0 n^{\prime} + \frac{1}{2} \mu \nu^{\prime} = 0 \ , \label{bgndfl_e}
\eeq  
where now
\beq
    {\cal B} = - 2 \frac{\partial \Lambda}{\partial n^2} \equiv 
               \frac{\mu}{n} \quad , \quad
    \b^0_0 = {\cal B} + 2 n^2 \frac{\partial {\cal B}}{\partial n^2} \ .
\eeq

There are several sets of ``boundary'' conditions that must be dealt with: 
at the center, at interfaces, and at the surface of the star.  \ In view of 
Eq.~(\ref{bckgrnd}), requiring a non-singular behavior at the center of the 
star will impose that $\lambda(0) = 0$, and consequently 
$\lambda^{\prime}(0)$ and $\nu^{\prime}(0)$ must also vanish. \ This in turn 
implies, in view of Eq.~(\ref{nprof}), that $n_\x^{\prime}(0)$ has to vanish 
as well. \ At the surface, we will only consider configurations that satisfy 
$n(R) = 0$. \ A smooth joining of the interior spacetime to a Schwarzschild 
vacuum exterior at the surface of the star implies that the total mass $M$ of 
the system is given by
\beq
    M = - 4 \pi \int^{R}_0 r^2~\Lambda(r) dr
\eeq
and that $\Psi(R) = 0$. \ The metric must be continuous across the interfaces, 
but the matter behaviour is a bit more complicated 
\cite{andersson02:_oscil_GR_superfl_NS}. \ We will discuss this in detail in 
Sec.~\ref{interface_join}.  

\subsection{The linearized field equations}

It is well-known that all non-trivial fluid pulsation modes of a nonrotating 
fluid star correspond to polar perturbations (often referred to as 
``even parity''). \ In the so-called Regge-Wheeler gauge 
\cite{regge57:_rw_gauge}, the corresponding metric components are
\beq
     \delta g_{\mu \nu} = - e^{i \omega t}
            \left[\begin{array}{cccc}
            e^{\nu}r^l H_{0}(r)&i \omega r^{l+1} H_{1}(r)&0&0 \cr 
            i \omega r^{l+1} H_{1}(r)&e^{\lambda} r^l H_{0}(r)&0&0 \cr 
            0&0&r^{l+2} K(r)&0\cr 
            0&0&0&r^{l+2} {\rm sin}^2\theta K(r)
            \end{array}\right] P_l(\theta) \ . \label{lemet}
\eeq
where $P_l(\theta)$ are the Legendre polynomials. \ This decomposition will 
be applied to each layer of the star.

The linearization of the fluid equations follows from the basic relation 
\bea 
    \delta \mu^\x_\mu &=& \left(\B{}_{\mu \nu} + {\cal A}^{\x \x}{}_{\mu \nu} 
    \right) \delta n^\nu_\x + \left({\cal X}^{\x \y}{}_{\mu \nu} + 
    \A{}_{\mu \nu}\right) \delta n^\nu_\y + \cr
    && \cr
    &&\frac{1}{2} g^{\tau \nu} \left(\delta^\sigma{}_\mu \mu^\x_\nu+ 
    \left[\B{}_{\mu \nu} + {\cal A}^{\x \x}{}_{\mu \nu}\right] n^\sigma_\x 
    + \left[{\cal X}^{\x \y}{}_{\mu \nu} + \A{}_{\mu \nu}\right] 
    n^\sigma_\y\right) \delta g_{\sigma \tau} \ , 
\eea 
where
\begin{eqnarray} 
    \B{}_{\mu \nu} &=& \B g_{\mu \nu} - 2 
    \frac{\partial \B}{\partial n^2_\x} g_{\mu \sigma} 
    g_{\nu \rho} n^\sigma_\x n^\rho_\x \ , \\
    && \cr
    {\cal X}^{\x \y}{}_{\mu \nu} &=& - 2 \frac{\partial \B}{\partial 
    n^2_\y} g_{\mu \sigma} g_{\nu \rho} n^\sigma_\x n^\rho_\y \ , \\
    && \cr
    {\cal A}^{\x \x}{}_{\mu \nu} &=& - g_{\mu \sigma} g_{\nu \rho} \left(
    \frac{\partial \B}{\partial n^2_{\x \y}} \left[n^\rho_\x 
    n^\sigma_\y + n^\sigma_\x n^\rho_\y\right] + \frac{\partial 
    \A}{\partial n^2_{\x \y}} n^\sigma_\y n^\rho_\y\right) \ , \\ 
                                && \cr
    \A{}_{\mu \nu} &=& \A g_{\mu \nu} - g_{\mu \sigma} 
    g_{\nu \rho} \left(\frac{\partial \B}{\partial n^2_{\x \y}} 
    n^\sigma_\x n^\rho_\x + \frac{\partial {\cal B}^\y}{\partial 
    n^2_{\x \y}} n^\sigma_\y n^\rho_\y + \frac{\partial \A} 
    {\partial n^2_{\x \y}} n^\sigma_\y n^\rho_\x\right) \ .  
\end{eqnarray}
 
These terms account for effects due to a single constituent's bulk ($\B$), 
the presence of multiple constituents ($\X$), and entrainment ($\A$). \ This
decomposition of the coefficients into these separate classes was first made 
by Andersson and Comer \cite{andersson07:_livrev}. \ Their relation to the 
coefficients mentioned earlier (which have been used in different 
applications  
\cite{comer99:_quasimodes_sf,andersson01:_slowl_rotat_GR_superfl_NS,andersson02:_oscil_GR_superfl_NS}) 
are
\bea
      \a^{\x \y}{}_0^0 &=& g^{0 0} \left({\cal X}^{\x \y}{}_{0 0} + 
                           \A{}_{0 0}\right) 
                   \ , \\
                && \cr
   \b^\x{}_0^0 &=& g^{0 0} \left(\B{}_{0 0} + {\cal A}^{\x \x}{}_{0 0}\right) 
                   \ . 
\eea

Writing the nucleon four-current as $n^\mu_\x = n_\x u^{\mu}_\x$, where 
$g_{\mu \nu} u^{\mu}_\x u^\nu_\x = - 1$, one finds that the velocity 
perturbation in the two-fluid layer is
\beq
    \delta u^i_\x = e^{- \nu/2} \frac{\partial}{\partial t} \delta \xi^i_\x 
                    \ ,
\eeq
where the displacement vector $\delta \xi^i_\x$ has components   
\beq
   \delta \xi_\x^r = e^{- \lambda/2} r^{l - 1} \wx(r) P_l e^{i \omega t} 
                     \quad , \quad
   \delta \xi_\x^{\theta} = - r^{l - 2} \vx(r) \frac{\partial}{\partial 
                            \theta} P_l e^{i \omega t} \ .  
\eeq
The Lagrangian variation for each nucleon number density can be written as
\begin{equation}
   \Delta n_\x = \delta n_\x + n^{\prime}_\x e^{- \lambda/2} 
                                r^{l - 1} \wx \ ,
\end{equation}
and the conservation equation Eq.~(\ref{conteq}) for each particle number 
current yields
\begin{equation}
   \frac{\Delta n_\x}{n_\x} = - r^l \left(e^{- \lambda/2} 
                \left[\frac{l + 1}{r^2} \wx + \frac{1}{r} W^{\prime}_\x
                \right] + \frac{l (l + 1)}{r^2} \vx - \frac{1}{2} H_0 - 
                K\right) \ . \label{lang}
\end{equation}
Thus, all matter variables can be expressed in terms of the velocity 
variables $\wx$ and $\vx$. \ In a single fluid layer, we would have as the 
only independent matter variables the two velocity components $W$ and $V$. 

The set of perturbation equations that we solve for in the superfluid layer 
have already been listed in \cite{comer99:_quasimodes_sf}, but since our 
multi-fluid and multi-layer problem requires a slightly different method of 
solution we repeat the relevant equations here. \ First we define for each 
species (in analogy with Lindblom and Detweiler's 
\cite{lindblom83:_quadoscs,detweiler85:_grpulse} approach to the one-fluid 
problem) the new variable 
\bea
    \xx &\equiv& n_\x \left[\frac{e^{\nu/2}}{2} \mu^\x \Ha + e^{-\nu/2} 
                 \omega^2 \left({\cal B}^\x n_\x \vx + \A n_\y V_\y\right)
                 \right] \cr 
              && \cr 
              && - e^{(\nu - \lambda)/2} \frac{n^{\prime}_\x}{r} \left(
                 \b^\x{}^0_0 n_\x \wx + \a^{\x \y}{}^0_0 n_\y W_\y\right) 
                 \ .  \label{xx}
\eea
Then we find that the Einstein and superfluid field equations yield an
algebraic constraint equation
\bea
  &&e^\lambda \left[\frac{2 - l - l^2}{r^2} - \frac{3}{r^2} \left(1 - 
    e^{- \lambda}\right) -8 \pi \Psi\right] \Ha + \left[\frac{2 \omega^2 
    }{e^{\nu}} - \frac{l (l + 1)}{2} e^\lambda 
    \left(\frac{1-e^{- \lambda}}{r^2} + 8 \pi \Psi\right)\right] \H1 \cr
  && \cr
  &&+ \left[-2 e^{\lambda - \nu} \omega^2 + e^\lambda \frac{l^2 + l - 2}{r^2} 
    + e^{2 \lambda} \left(\frac{1 - e^{- \lambda}}{r^2} + 8 \pi \Psi
    \right)\left(1 - \frac{3}{2} \left(1 - e^{- \lambda}\right) - 4 \pi 
    r^2 \Psi\right)\right] K \cr
  && \cr 
  &&+ 16 \pi e^{\lambda - \nu/2} \left(X_\n + X_\p\right) = 0 \ , 
    \label{cnstrt}
\eea
and a system of coupled ordinary differential equations [where we use the 
definition $\d^0_0 = \b^\n{}^0_0 \b^\p{}^0_0 - (\a^{\n \p}{}^0_0)^2$]:
\bea
    \H1^{\prime} &=& \frac{e^\lambda}{r} \Ha + \left(\frac{\lambda' - \nu'}{2} 
    - \frac{l + 1}{r}\right) \H1 + \frac{e^\lambda}{r} K - 16 \pi 
    \frac{e^\lambda}{r} \left(\mu^\n n_\n V_\n + \mu^\p n_\p V_\p\right) 
    \ , \label{h1prime} \\
  && \cr
    K^{\prime} &=& \frac{\Ha}{r} + \frac{l (l + 1)}{2 r} \H1 + 
    \left(\frac{\nu^{\prime}}{2} - \frac{l + 1}{r}\right) K - 8 \pi 
    \frac{e^{\lambda/2}}{r} \left(\mu^\n n_\n W_\n + \mu^\p n_\p W_\p\right) 
    \ , \\
  && \cr
    \wx^{\prime} &=& \frac{e^{\lambda/2} r}{2} \Ha + e^{\lambda/2} r K - 
    e^{\lambda/2} \frac{l (l + 1)}{r} \vx - \left(\frac{l + 1}{r} + 
    \frac{n^{\prime}_\x}{n_\x}\right) \wx \cr
  && \cr
  &&+ \frac{\b^\y{}^0_0}{n_\x^2 \d^0_0} \left[e^{(\lambda - \nu)/2} r \xx + 
    n^{\prime}_\x \left(\b^\x{}^0_0 n_\x \wx + \a^{\x \y}{}^0_0 n_\y W_\y
    \right)\right] \cr
  && \cr
  &&- \frac{\a^{\x \y}{}^0_0}{n_\x n_\y \d^0_0} \left[e^{(\lambda - \nu)/2} r 
    X_\y + n_\y^{\prime} \left(\a^{\x \y}{}^0_0 n_\x \wx + \b^\y{}^0_0 n_\y 
    W_\y\right)\right] \ , \label{Wnprim} \\
  && \cr
    \xx^{\prime} &=& - \frac{l}{r} \xx + \frac{e^{\nu/2}}{2} \left[\mu^\x n_\x
    \left(\frac{1}{r} - \nu^{\prime}\right) - n_\x^{\prime} \left(\b^\x{}^0_0 
    n_\x + \a^{\x \y}{}^0_0 n_\y\right)\right] \Ha \cr
  && \cr
  &&+ \mu^\x n_\x \left[\frac{e^{\nu/2}}{4} \frac{l (l + 1)}{r} + 
    \frac{\omega^2}{2} r e^{-\nu/2}\right] \H1 \cr
  && \cr
  &&+ e^{\nu/2}\left[\mu^\x n_\x \left(\frac{\nu^{\prime}}{4} - \frac{1}{2 r}
    \right) - \left(\b^\x{}^0_0 n_\x + \a^{\x \y}{}^0_0 n_\y\right) 
    n^{\prime}_\x\right] K \cr
  && \cr
  &&+ \frac{l (l + 1)}{r^2} e^{\nu/2} n^{\prime}_\x \left(\b^\x{}^0_0 n_\x \vx 
    + \a^{\x \y}{}^0_0 n_\y V_\y\right) 
    - e^{(\lambda - \nu)/2} \frac{\omega^2}{r} n_\x \left(\B n_\x \wx + \A 
    n_\y W_\y\right) \cr
  && \cr
  &&- 4 \pi e^{(\lambda + \nu)/2} \frac{\mu^\x n_\x}{r} 
    \left(\mu^\x n_\x \wx + \mu^\y n_\y W_\y\right) 
    + e^{-(\lambda - \nu)/2} \left[- \frac{n^{\prime}_\x}{r} \left(
    \b^\x{}^0_0{}^{\prime} n_\x \wx \right. \right. \cr
  && \cr
  &&\left. \left.+ \a^{\x \y}{}^0_0{}^{\prime} n_\y 
    W_\y\right) + \left(\frac{2 n^{\prime}_\x}{r^2} + \frac{\lambda^{\prime} - 
    \nu^{\prime}}{2 r} n^{\prime}_\x - \frac{n^{\prime \prime}_\x}{r}\right)
    \left(\b^\x{}^0_0 n_\x \wx + \a^{\x \y}{}^0_0 n_\y W_\y\right)\right] \ . 
    \label{xpprime}
\eea

In the one-fluid case, the constraint equation becomes
\bea
   &&e^\lambda \left[\frac{2 - l -l^2}{r^2} - \frac{3}{r^2} \left(1 - 
     e^{- \lambda}\right) - 8 \pi \Psi\right] \Ha + \left[
     \frac{2 \omega^2}{e^{\nu}} - \frac{l (l + 1)}{2} e^\lambda \left(
     \frac{1 - e^{- \lambda}}{r^2} + 8 \pi \Psi\right)\right] \H1 \cr
   && \cr
   &&+ \left\{- 2 e^{\lambda - \nu} \omega^2 + e^\lambda 
     \frac{l^2 + l - 2}{r^2} + e^{2\lambda} \left[
     \frac{1 - e^{- \lambda}}{r^2} + 8 \pi \Psi\right] \left[1 - \frac{3}{2} 
     \left(1 - e^{- \lambda}\right) - 4 \pi r^2 \Psi\right]\right\} K \cr
   && \cr
   &&+ 16 \pi e^{\lambda - \nu/2} X = 0 \ ,
   \label{envcon}
\eea
and the other two equations for the metric are
\bea
   \H1^{\prime} &=& \frac{e^\lambda}{r} \Ha + \left[\frac{\lambda' - \nu'}{2} 
                    - \frac{l + 1}{r}\right] \H1 + \frac{e^\lambda}{r} K - 16 
                    \pi \frac{e^\lambda}{r} \mu n V \ , \label{h1env} \cr 
                 && \cr
     K^{\prime} &=& \frac{\Ha}{r} + \frac{l (l+1)}{2 r} \H1 + \left[
                  \frac{\nu^{\prime}}{2} - \frac{l + 1}{r}\right] K - 
                  8 \pi \frac{e^{\lambda/2}}{r} \mu n W \ . 
                  \label{kenv}
\eea
The final two fluid equations are 
\bea
   W^{\prime} &=& \frac{e^{\lambda/2} r}{2} \Ha + e^{\lambda/2} r K - 
                  e^{\lambda/2} \frac{l (l + 1)}{r} V - \frac{l + 1}{r} W + 
                  \frac{e^{(\lambda-\nu)/2} r}{n^2 {\rm B}^0_0} X \ , 
                  \label{wpenv} \cr
               && \cr
   X^{\prime} &=& - \frac{l}{r} X + \frac{e^{\nu/2}}{2} \left[n \mu 
                  \left(\frac{1}{r} - \nu^{\prime}\right) - n^{\prime} 
                  {\rm B}^0_0 n\right] \Ha + \mu n \left[\frac{e^{\nu/2}}{4} 
                  \frac{l (l + 1)}{r} + \frac{\omega^2}{2} r e^{-\nu/2}
                  \right] \H1 \cr
               && \cr
               && + e^{\nu/2} \left[\mu n \left(\frac{\nu^{\prime}}{4} 
                  - \frac{1}{2 r}\right) - {\rm B}^0_0 n n^{\prime}\right] K 
                  + \frac{l (l + 1)}{r^2} e^{\nu/2} n^{\prime} {\rm B}^0_0 n 
                  V \cr
               && \cr
               && - e^{(\lambda-\nu)/2} \frac{\omega^2}{r} {\cal B} n^2 W - 4 
                  \pi e^{(\lambda+\nu)/2} \frac{(\mu n)^2}{r} W \cr
               && \cr
               && + e^{-(\lambda-\nu)/2} \left[- \frac{n^{\prime}}{r} 
                  {\rm B}^0_0{}^{\prime} n W + \left(
                  \frac{2 \p^{\prime}}{r^2} + 
                  \frac{\lambda^{\prime}-\nu^{\prime}}{2 r} n^{\prime} - 
                  \frac{n^{\prime \prime}}{r}\right) {\rm B}^0_0 n W\right] 
                  \ , \label{xpenv}
\eea
where
\beq
    X = n \left[\frac{e^{\nu/2}}{2} \mu \Ha + e^{- \nu/2} \omega^2 
          \left({\cal B} n V\right)\right] - e^{(\nu-\lambda)/2} 
          \frac{n^{\prime}}{r} \left({\rm B}^0_0 n W\right) \ .
\eeq

Outside the star, the problem is reduced to solving the so-called Zerilli 
equation. \ We refer the reader to \cite{comer99:_quasimodes_sf} for details 
on how to match the interior and exterior solutions.

\subsection{Interface Layer Junction Conditions} \label{interface_join}

At the center of the star, the boundary conditions are those given in 
Appendix A of Comer et al \cite{comer99:_quasimodes_sf}, i.e.~all functions 
are regular. \ At the surface, the conditions are the one-fluid conditions 
used by Detweiler and Lindblom 
\cite{lindblom83:_quadoscs,detweiler85:_grpulse}.  \ The main difference here 
concerns the interfaces between the single-fluid and two-fluid layers. \ The 
detailed treatment of an interface was discussed by Andersson et al 
\cite{andersson02:_oscil_GR_superfl_NS}. \ They found that the relativistic 
junction conditions imply that the three metric perturbations $H_0, H_1$ and 
$K$ must be continuous at each interface. 

To make progress we also need to specify the behaviour of the superfluid 
neutron velocity, in practice $W_\n$, at each interface. \ A truly realistic 
model would require a detailed analysis of the phase-transition from the 
single fluid regime to the two-fluid domain. \ The physics of chemical 
equilibrium is more complicated than that of strict particle number density 
current conservation. \ Hence the interface behaviour is likely to be 
complex, with finite temperature effects playing a crucial role. \ To see 
that this must be the case is easy. 

The superfluid energy gap $\Delta$ varies with the density, 
cf.~Fig.~\ref{back-dens}, and for any given temperature $T$ one would expect 
the transition to superfluidity to take place when $k_B T \approx \Delta$. \ 
In a region of ``strong superfluidity'', where $k_\b T \ll \Delta$, one can 
safely ignore finite temperature effects. \ However, as one approaches the 
phase transition (at points where $k_B T \to \Delta$) excitations will play 
an increasing role. \ In fact, they dictate how the two distinct fluid 
degrees of freedom couple as one transits into the single-fluid regime. \ 
This transition problem has not yet been investigated in detail. \ Hence, we 
must make simplifications in order to proceed. 

One can think of two natural limits. \ In the first, the particle reactions 
that drive the system towards chemical equilibrium are slow compared to the 
mode oscillation in the transition region, leading to a situation where a 
``superfluid'' fluid element can temporarily move into the single fluid 
region without losing its identity. \ In that case, $W_\n$ would be freely 
specificable at the interface. \ In the opposite limit, when reactions are 
faster than the timescale of oscillation, a fluid element that moves across 
the interface would immediately lose its identity. \ This would lock the 
velocities at the interface, and as a result $W_\n$ would be linked to the 
single-fluid velocity $W$. \ In our numerical calculations we assume that it 
is this latter scenario that applies. \ This is in contrast with 
\cite{andersson02:_oscil_GR_superfl_NS}, where it was assumed that  reactions 
are slow.

\section{Analytic expansion for the local matter content} \label{anexp}

The perturbation equations require as input information about the bulk, 
multi-constituent, and entrainment effects. \ This necessarily leads to a 
number of coefficients to be determined. \ Admittedly, for those not familiar 
with the multi-fluid formalism (some of) these coefficients may seem somewhat 
obscure. \ In particular, since most of them tend to be ignored in 
discussions of the supranuclear equation of state. \ However, for our present 
purposes, the study of neutron star quasinormal modes, we can simplify the 
problem. \ To do this we consider an expansion of the master function that 
accounts for the fact that our background configuration is static and 
spherically symmetric. \ Alternatively, the expansion can be interpreted as 
one where the fluid velocities are small compared to the speed of light 
\cite{andersson02:_oscil_GR_superfl_NS,comer03:_rel_ent}. \ For mode 
oscillations, this is a reasonable assumption. \ It is also expected to be 
accurate for slow-rotation models, see 
\cite{andersson01:_slowl_rotat_GR_superfl_NS}.

Consider, for example, a term like $\partial \A / \partial n^2_{\n \p}$. \ In 
principle, it implies that we need to expand the master function to 
${\cal O}(n^4_{\n \p})$. \ On the scale of a fluid element spacetime can be 
taken to be that of Minkowski. \ The $n^2_{\n \p}$ term can then be written 
as  
\beq
    n^2_{\n \p} = n_\n n_\p \left(\frac{1 - \vec{v}_{\n} \cdot 
                  \vec{v}_{\p}/c^2} {\sqrt{1 - (v_{\n}/c)^2} 
                  \sqrt{1 - (v_{\p}/c)^2}}\right) \ ,
\eeq
where $\vec{v}_\n$ and $\vec{v}_\p$ are, respectively, the neutron and proton 
three-velocities in the local Minkowski frame. \ When the individual 
three-velocities $\vec{v}_{\n ,\p}$ satisfy $v_{\n,\p} / c \ll 1$, 
then $n^2_{\n \p} = n_\n n_\p$ up to first-order in the ratio $v_\x/c$. 

With this as our guide we write the master function in the form 
\cite{andersson02:_oscil_GR_superfl_NS,comer03:_rel_ent}
\beq
    \Lambda(n^2_\n,n^2_\p,n^2_{\n \p}) = \sum_{i = 0}^{\infty} 
            \lambda_i(n^2_\n,n^2_\p) \left(n^2_{\n \p} - n_\n n_\p\right)^i \ .
\eeq
The bulk, multi-constituent, and entrainment coefficients thus become  
\begin{eqnarray}
  {\cal A}^{\n \p} &=& - \lambda_1 - \sum_{i = 2}^{\infty} i~\lambda_i 
                       \left(n^2_{\n \p} - n_\n n_\p\right)^{i - 1}  \ , \\
     && \cr
  \B &=& - \frac{1}{n_\x} \frac{\partial \lambda_0}{\partial n_\x} - 
         \frac{n_\y}{n_\x} \A - \frac{1}{n_\x} \sum_{i = 1}^{\infty} 
         \frac{\partial \lambda_i}{\partial n_\x} \left(n^2_{\n \p} - 
         n_\n n_\p\right)^i \ , \\
     && \cr
  \a^{\x \y}{}^0_0 &=& - \frac{\partial^2 \lambda_0}{\partial n_\x \partial 
              n_\y} 
         - \sum_{i = 1}^{\infty} \frac{\partial^2 \lambda_i}{\partial n_\x 
         \partial n_\y} \left(n^2_{\n \p} - n_\n n_\p\right)^i \ , \\
     && \cr
  \b^\x{}^0_0 &=& - \frac{\partial^2 \lambda_0}{\partial n^2_\x} - 
           \sum_{i = 1}^{\infty} \frac{\partial^2 \lambda_i}{\partial 
           n^2_\x} \left(n^2_{\n \p} - n_\n n_\p\right)^i \ , \\
     && \cr
  \frac{\partial \A}{\partial n_\x} &=& - \frac{\partial \lambda_1}
           {\partial n_\x} - \sum_{i = 2}^{\infty} i \left(\frac{\partial 
           \lambda_i}{\partial n_\x} \left[n^2_{\n \p} - n_\n n_\p\right] - 
           \left[i - 1\right] n_\y \lambda_i\right) \left(n^2_{\n \p} - n_\n 
           n_\p\right)^{i - 2} \ , \\   
        && \cr
  \frac{\partial \A}{\partial n^2_{\n \p}} &=& - 2 \lambda_2 - 
           \sum_{i = 3}^{\infty} i \left(i - 1\right) \lambda_i 
           \left(n^2_{\n \p} - n_\n n_\p\right)^{i - 2} \ .
\end{eqnarray}

The $\A$ and $\B$ coefficients on the background are
\beq
    \A = - \lambda_1 \quad , \quad \B = - \frac{1}{n_\x} 
         \frac{\partial \lambda_0}{\partial n_\x} + \frac{n_\y}{n_\x} 
         \lambda_1 \ .
\eeq
This implies for the background chemical potentials that
\beq
    \mu^\x = - \frac{\partial \lambda_0}{\partial n_\x} \ .
\eeq
In the single-fluid case, $- \lambda_0$ is the energy density, and so the 
chemical potential above is equal to that of the single-fluid result. \ We 
now see that the other coefficients simplify to
\beq
    \a^{\x \y}{}^0_0 = \frac{\partial \mu^\x}{\partial n_\y} = 
                  \frac{\partial \mu^\y}{\partial n_\x} \quad , \quad 
    \b^\x{}^0_0 = \frac{\partial \mu^\x}{\partial n_\x} \ .
\eeq

To gain a little further insight, a local, plane wave analysis 
\cite{andersson07:_livrev} indicates that the sound speeds (plural, because 
there are two) are the solutions to
\beq
    \left({\cal B}^\n c^2_s - \b^\n{}^0_0\right) \left({\cal B}^\p 
    c^2_s - \b^\p{}^0_0\right) + \left({\cal A}^{\n \p} c^2_s - 
    \a^{\n \p}{}^0_0\right)^2 = 0 \ . \label{sndspd}
\eeq
We see that in addition to the expected bulk contribution to the sound speed 
(through ${\cal B}^\x$ and $\b^\x{}^0_0$) we have also entrainment entering 
via ${\cal A}^{\n \p}$ and the ``symmetry energy'' through 
$\a^{\n \p}{}^0_0$. \ These effects at the local level have global 
consequences. \ For example, Andersson et al 
\cite{andersson02:_oscil_GR_superfl_NS} have shown that entrainment has an 
impact on the oscillation mode spectrum by introducing so-called 
``avoided crossings'' between the ordinary and superfluid modes as the 
entrainment is varied. \ Prix et al \cite{prix02:_slow_rot_ns_entrain} have 
demonstrated that the symmetry energy leaves its imprint on slowly rotating 
configurations when the neutrons and protons are not co-rotating.

Another dramatic effect of the entrainment is that it may facilitate a 
two-stream instability \cite{acp03:_twostream_prl,andersson04:_twostream}. \ 
In principle, such instabilities may operate in any system with two 
inter-penetrating components moving at different speeds. \ The key 
requirement for the instability to act is that the the two fluids are 
coupled, and entrainment can facilitate this coupling. \ The superfluid 
two-stream instability is being considered as a possible trigger for glitches 
and appears to be consistent with seven years of glitch data from the fastest 
known young pulsar J0537 \cite{middleditch06:_glitches}. 

\section{Equation of State Input} \label{masfcn}

\subsection{Master Function Modulo Entrainment}

In our numerical calculations we will compare two relatively simple
parametrisations for the equation of state. \ Nothing prevents us from 
considering other perhaps more realistic models, including tabulated equation 
of state data, but at this point we are more concerned with the technical 
developments than with the actual numerical results. \ As we will discuss 
below, see Sec.~\ref{neweos}, a fully consistent model requires information 
that is not yet available. \ Until this changes, this kind of analysis must 
be viewed as somewhat qualitative.  

We consider two models for $\lambda_0$: i) a simple ``TOY'' model that is a 
sum of polytropes for the neutrons and protons, a symmetry energy term, and a 
final term that accounts for a relativistic gas of electrons, and ii) the 
more realistic, so-called ``PAL'' equation of state \cite{pal88:_PhRvL}.  

Specifically, the TOY model in the one-fluid and two-fluid layers takes the 
form
\beq
    \lambda_0/m = - \left(n_\n + n_\p\right) - \sigma_{\rm b} 
                  \left(n_\n + n_\p\right)^{\beta_{\rm b}} - S_0 
                  \left(n_\n + n_\p\right) \left(1 - 2 x_\p\right)^2 - 
                   \sigma_\mathrm{e} n_\p^{\beta_\mathrm{e}} \ ,
\eeq
where we have chosen the parameters $\sigma_{\rm b} = 0.2$, 
$\sigma_\mathrm{e} = 0.5$, $\beta_{\rm b} = 2.5$, $\beta_\mathrm{e} = 2.0$, 
$S_0 = 0.05$, $m$ is the baryon mass, and $x_\p = n_\p/n$ is the proton 
fraction. \ As shown in Table~\ref{ptable}, this combination leads to a 
neutron star model with reasonable mass and radius. 

The particular PAL master function for both layers is 
\cite{andersson04:_twostream} 
\beq
    \lambda_0/m = - n_0 u \left[1 + E_0(u) + S(u) 
                  \left(1 - 2 x_\p\right)^2\right] - \lambda_\mathrm{e}/m \ ,
\eeq
where
\bea
     m E_0(u) &=& A_0 u^{2/3} + B_0 u + C_0 u^\sigma + 3 \sum_{i = 1}^2 C_i 
               \alpha_i^{- 3} \left[\alpha_i u^{1/3} - \tan^{- 1} 
               \left(\alpha_i u^{1/3}\right)\right] , \cr
            && \cr
     m S(u) &=& A_s \left(u^{2/3} - u\right) + S_0 u \ , \cr
              && \cr 
    \lambda_e &=& \frac{m_\e}{\tau^3_\e} \chi(\chi^F_\e) \ , \cr
               && \cr
    \chi(x) &=& \frac{1}{8 \pi^2} \left\{x \left(1 + x^2\right)^{1/2} 
                \left(1 + 2 x^2\right) - 
                \ln \left[x + \left(1 + x^2\right)^{1/2}\right]\right\} \ , \cr
             && \cr 
    \chi^F_\e &=& 1836 \left[3 \pi^2 
                  \left(\frac{\hbar}{m}\right)^3\right]^{1/3} n^{1/3}_\p 
                  \quad \ , \quad \tau_e = \hbar/m_\e \ ,
\eea
with
$u = n/n_0$ ($n_0 = 0.16~{\rm fm}^{-3}$), $\sigma = 0.927$, 
$A_0 = 22.11~{\rm MeV}$, $B_0 = 220.47~{\rm MeV}$, $C_0 = - 213.41~{\rm MeV}$, 
$C_1 = - 83.84~{\rm MeV}$, $C_2 = 23.0~{\rm MeV}$, $\alpha_1 = 2/3$, 
$\alpha_2 = 1/3$, $A_s = 12.99~{\rm MeV}$, and $S_0 = 30~{\rm MeV}$.
Again, the values for mass and radius obtained for these parameter
values are reasonable, see Table~\ref{ptable}.

\begin{table}
 \begin{tabular}[t]{|c|c|c|c|c|c|}
 \hline 
 Model & $~T~({\rm MeV})~$ & $~n(0)~({\rm fm}^{- 3})~$ & $~M~(M_{\odot})~$ & 
 $~R~({\rm km})~$ \\
 \hline
 TOY & ---  & 1.0 & 1.635 & 10.655 \\
 \hline
     & 0.45 & 1.0 & 1.745 & 10.641 \\
 PAL & 0.5 & 1.0 & 1.751 & 10.661 \\
     & 0.55 & 1.0 & 1.759 & 10.684 \\
 \hline
 \end{tabular}
\caption{The ``canonical'' background stellar models for the TOY and PAL 
equations of state for the parameter values discussed in the main text.} 
\label{ptable}
\end{table}

The electron contribution $\lambda_\e$ is vital for ensuring chemical 
equilibrium of the system. \ Even though the electron mass $m_\e$ is much 
smaller than the nucleon mass ($m_\e = m/1836$), the fact that the electrons 
are ultra-relativistic gives them enough energy to affect the chemical 
potential at the same level as the nucleons. \ Because of overall charge 
neutrality we set $n_\e = n_\p$, and recall that the electrons and the 
protons flow together. \ Note also that, even though the imposition of 
chemical equilibrium will effectively make the single-fluid and two-fluid 
equations of state the same on the background, we have to distinguish them 
because of the $\B$, $\b^\x{}^0_0$, etc.~coefficients, for which the partial 
derivatives must be computed before chemical equilibrium is imposed.

\subsection{Gap Model} \label{gapsec}

BCS theory is the basic paradigm for Fermionic superfluidity. \ Given a 
many-particle system with an attractive component in the interactions, the 
theory tells us that Cooper pairs will form, leading to a fundamental 
modification of the energy states near the Fermi surface, namely the 
formation of a gap in the energy spectrum. \ It is the existence of this gap 
that leads to superfluidity and in the present context multi-fluid dynamics. 
\ Even if particles try to scatter dissipatively, there are essentially no 
accessible states for them to scatter into unless there is enough energy to 
break a Cooper pair. \ When pair-breaking occurs the particles can reach the 
states above the gap, and energy can be irreversibly dissipated. \ Those 
particles in energy states ``below'' the Fermi surface are in this sense also 
``superfluid'' since they cannot enter already filled states. \ So, for 
example, when neutrons are superfluid, ordinary scattering that would cause  
(on average) neutrons and protons to flow as a single fluid is severely 
diminished. \ It is thus clear that neutron star dynamics on a macroscopic 
scale is dictated by the gap structure on the microscopic scale.

There has been much work analyzing the details of the gap structure in dense, 
nucleonic matter trying to account for different interactions, medium effects 
etcetera (see \cite{lombardo01:_lnp} for a review of gaps in general 
and \cite{andersson04:_viscosity_coeff} for an application to neutron stars). 
\ If even our rudimentary understanding of the strong force is correct for 
the densities expected in neutron stars there is little doubt that neutron 
and proton gaps exist. \ But given the complexities of the problem there is 
no exact agreement on the details; particularly on the density dependence of 
the gap. \ This affects the maximum gap energy, the gap profile as a function 
of density, and the size of regions in which gaps exist. \ However, it is 
generally accepted that neutron superfluidity and proton superconductivity 
will not extend throughout the star. \ Consequently, there should be layers 
having ordinary, single-fluid dynamics as well as ones where the multi-fluid 
description applies. \ This implies that most, if not all, previous mode 
calculations and modeling of rotational equilibria that assume superfluidity 
throughout the star are incomplete (see 
\cite{comer02:_zero_freq_subspace,andersson07:_livrev} for reviews). 

As a first step towards improving the situation we will consider models that 
account for the detailed energy gap and its dependence on, in particular, the 
density. \ For practical reasons, we will use the parametrized model of 
\cite{kaminker01:_super_gap,yakolev01:_super_cool,kaminker02:_super_cool,yakovlev02:_conf} 
as adapted by Andersson et al \cite{andersson04:_viscosity_coeff}. \ This 
model has a number of ``free'' parameters, that can be used, for instance, to 
adjust the maximum gap energy $\Delta_0$ and the gap profile (as a function 
of density). \ In this phenomenological description, the gap energy 
$\Delta(k_F)$ (where $k_F$ is the wave-number at the Fermi surface) takes the 
form
\beq
     \Delta(k_F) = \Delta_0 \frac{\left(k_F - k_1\right)^2}
                   {\left(k_F - k_1\right)^2 + k_2} 
                   \frac{\left(k_F - k_3\right)^2}
                   {\left(k_F - k_3\right)^2 + k_4} \ . \label{gap}
\eeq
Note that density dependence enters implicitly through the wave-number $k_F$. 
\ Andersson et al \cite{andersson04:_viscosity_coeff} provide a table of 
parameter values representative of the various gap calculations in the extant 
literature. \ For our numerical calculations we focus on gap model ``h'' 
given in their Table~1.

Once an equilibrium configuration is built, and density is known as a 
function of radius, we can use the gap energy to determine which layers of 
the equilibrium configuration are ordinary or superfluid. \ This requires an 
assumption of the temperature profile in the star. \ In principle, our 
calculations assume that the fluid is at zero temperature, somewhat in the 
same spirit as mode-damping calculations due to shear and/or bulk viscosity 
\cite{na03:_rev}. \ We certainly do not have a consistent temperature 
description. \ However, for our present purposes, this is not a major problem. 
\ The gap information that we use is phenomenological so it should be 
acceptable to account for the temperature in an approximate way as well. \ In 
view of this, we simply assume that the fluid is isothermal (and do not 
account for the gravitational redshift). 

It would certainly be possible to improve on this description and it will 
eventually be important to do so. \ However, one would then like to account 
for all temperature effects, including quasiparticle excitations in the 
superfluid. \ This is an interesting problem since it opens the door for 
studies of dissipative superfluid neutron stars, but we will not discuss it 
further here. 

\begin{figure}[t]
\centering
\includegraphics[width=10cm,clip]{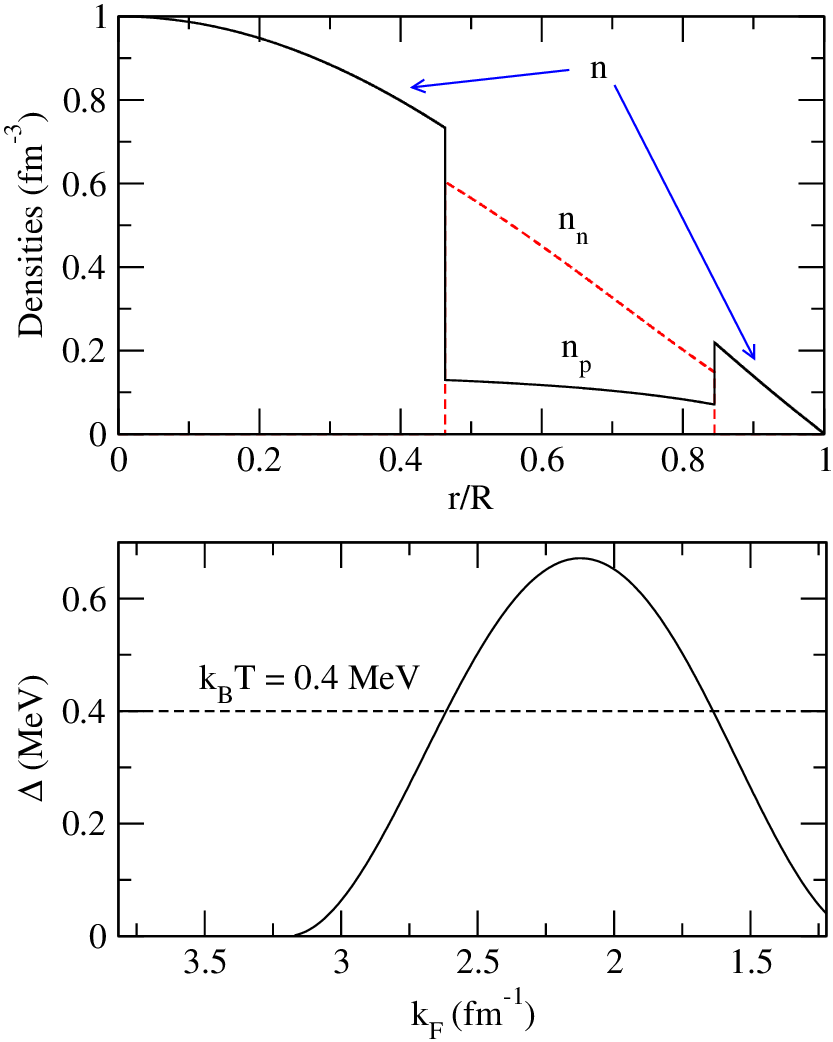}
\caption{Plot of the TOY model density profiles (vs radius) and gap function 
(vs the Fermi momentum) for the case $k_B T = 0.4$ MeV. \ The dashed line is 
the superfluid neutron density, and in the same range the solid line is that 
of the protons. \ Otherwise, the solid line represents the total baryon number 
density in the non-superfluid regions. \ If the gap energy is higher than the 
thermal energy, the neutrons are taken to be superfluid.}
\label{back-dens}
\end{figure}

Once we have chosen a core temperature we can work out if there are 
superfluid regions in the star. \ Varying the temperature thus affects the 
size of the superfluid layer, but has only a small effect on global parameters 
like the mass and radius, see Table~\ref{ptable}. \ We also adapt the number 
of independent fluids for the mode calculations accordingly. \ For our 
canonical TOY and PAL neutron star models, the particle number density and 
gap energy as a function of radius typically take the form shown in 
Fig.~\ref{back-dens}. \ As anticipated, there is a layering of regions having 
different fluid dynamics. \ Wherever the gap energy $\Delta$ is greater than 
the thermal energy $k_B T$, the neutrons will be superfluid, and therefore 
two-fluid dynamics will apply.

\subsection{The $\sigma$ - $\omega$ Relativistic Mean Field Model} 
\label{meanfield}

The model to be used for entrainment has been obtained using a relativistic
$\sigma - \omega$ mean field model of the type described by Glendenning 
\cite{glendenning97:_compact_stars}. \ Whatever reservations one may have 
about mean-field equations of state, the great advantage from our present 
perspective is that the entrainment can be quantified. \ The Lagrangian for 
this system is given by
\beq
    L = L_{b} + L_{\sigma} + L_{\omega} + L_{int} \ ,
\eeq
where
\begin{eqnarray}
   L_{b} &=& \bar{\psi} (i\gamma _{\mu } \partial^{\mu } - m) \psi \ ,
             \\
          && \cr
   L_{\sigma} &=& - \frac{1}{2} \partial _{\mu} \sigma \partial^{\mu}
                   \sigma -\frac{1}{2}m_{\sigma }^{2} \sigma ^{2} \ , \\
               && \cr
   L_{\omega} &=& - \frac{1}{4} \omega _{\mu \nu} \omega ^{\mu \nu}
                  - \frac{1}{2} m_{\omega}^{2} \omega _{\mu} 
                  \omega ^{\mu} \ , \\
               && \cr
   L_{int} &=& g_{\sigma} \sigma \bar{\psi} \psi - g_{\omega} 
               \omega _{\mu } \bar{\psi} \gamma^{\mu }\psi \ .
\end{eqnarray}
Here $m$ is the baryon mass, $\psi$ is an 8-component spinor with the proton 
components as the top 4 and the neutron components as the bottom 4, the 
$\gamma _{\mu}$ are the corresponding $8 \times 8$ block diagonal Dirac 
matrices, and $\omega _{\mu \nu} = \partial _{\mu} \omega _{\nu} - 
\partial_{\nu} \omega _{\mu }$. \ The coupled set of field equations obtained 
from this Lagrangian are to be solved in each fluid element. 

The main approximations of the mean field approach are to assume that the 
nucleons can be represented as plane-wave states and that all gradients of 
the $\sigma$ and $\omega^\mu$ fields can be ignored. \ The coupling constants 
$g_\sigma$ and $g_\omega$ and field masses $m_\sigma$ and $m_\omega$ are 
determined, for instance, from properties of nuclear matter at the nuclear 
saturation density. \ Fortunately, in what follows, we only need to provide 
the ratios $c^2_\sigma = (g_\sigma/m_\sigma)^2$ and $c^2_\omega = 
(g_\omega/m_\omega)^2$.

The main aim here is to produce a master function that incorporates the 
entrainment effect. \ Consider again fluid elements somewhere in the neutron 
star. \ The fermionic nature of the nucleons means that they are to be placed 
into the various energy levels (obtained from the mean field calculation) 
until their respective (local) Fermi spheres are filled. \ The Fermi spheres 
are surfaces in momentum space. \ The entrainment is incorporated by 
displacing the center of the proton Fermi sphere from that of the neutron 
Fermi sphere. \ The neutron sphere is centered on the origin, and has a 
radius $k_\n$. \ Displaced an amount $K$ from the origin is the center of the 
proton sphere, which has a radius $k_\p$. \ The Fermi sphere radii and 
displacement $(k_\n,k_\p,K)$ are functions of the local neutron and proton 
number densities and the local, relative velocity of the protons, say, with 
respect to the neutrons. 

Earlier we discussed the analytic expansion that applies to the master 
function, and determined that we need to know the $\lambda_0$ and $\lambda_1$ 
coefficients. \ The mean field model yields \cite{comer03:_rel_ent} 
\begin{eqnarray}
   \lambda_1 &=& - c_{\omega}^2 - \frac{c^2_{\omega}}{5 \mu^\n{}^2} 
                 \left[2 k^2_\p \frac{\sqrt{k^2_\n + \ms^2}} 
                 {\sqrt{k^2_\p + \ms^2}} + \frac{c^2_{\omega}}{3 \pi^2} 
                 \left(\frac{k^2_\n k^3_\p}{\sqrt{k^2_\n + \ms^2}} + 
                 \frac{k^2_\p k^3_\n}{\sqrt{k^2_\p + \ms^2}}\right)
                 \right] - \cr
              && \cr
              && \frac{3 \pi^2 k^2_{\p}}{5 \mu^\n{}^2 k^3_\n} 
                 \frac{k^2_\n + \ms^2}{\sqrt{k^2_\p + \ms^2}} \ ,
\end{eqnarray}
where the Dirac effective mass $\ms$ is the solution to the transcendental 
equation 
\begin{eqnarray}
  \ms &=& m - \ms \frac{c^2_{\sigma}}{2 \pi^2} \left[ k_\n \sqrt{k_\n^2
            + \ms^2} + k_\p \sqrt{k_\p^2 + \ms^2} - \right.\cr
        && \cr
        && \left.\frac{1}{2} \ms^2 {\rm ln} \left(\frac{k_\n + 
           \sqrt{k_\n^2  + \ms^2}}{- k_\n + \sqrt{k_\n^2 + \ms^2}}
           \right) - \frac{1}{2} \ms^2 {\rm ln} \left(\frac{k_\p + 
           \sqrt{k_\p^2 + \ms^2}}{- k_\p + \sqrt{k_\p^2 + \ms^2}}
           \right)\right] \ , \label{diracmass}
\end{eqnarray}
and each nucleon number density $n_\x$ is related to its Fermi surface radius 
$k_\x$ via
\beq
    n_\x = \frac{k^3_\x}{3 \pi^2} \ .
\eeq

\begin{figure}[t]
\centering
\includegraphics[height=5cm,clip]{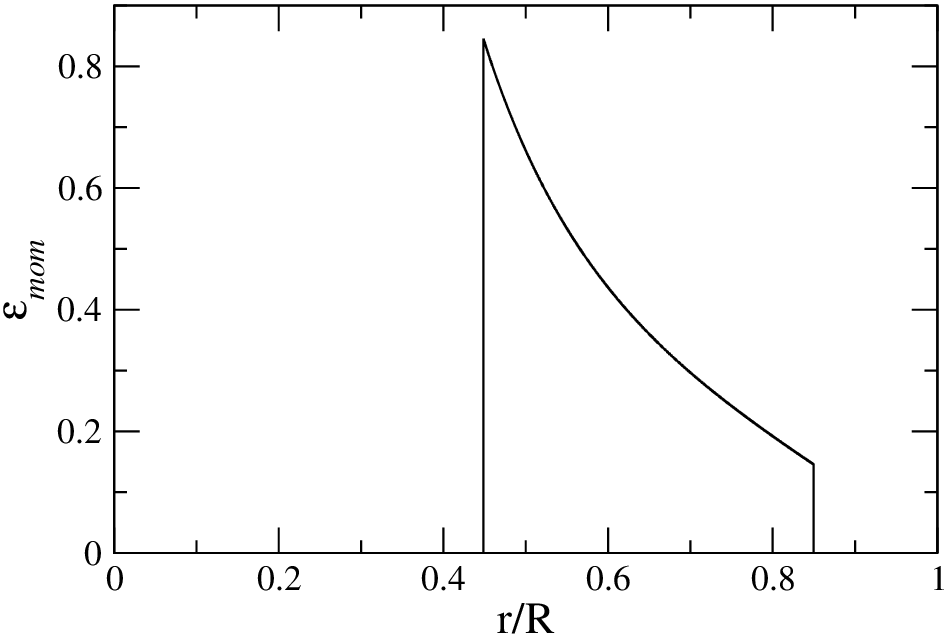}
\caption{The radial profile of $\varepsilon_{mom}$ defined in 
Eq.~(\ref{epsmom}) (for $c_\omega^2 = 7.148$ and $c_\sigma^2 = 12.684$) using 
the same TOY model as in Fig.~\ref{back-dens}.}
\label{ent_param}
\end{figure}

Typical behaviour of the entrainment parameter $\varepsilon_{mom}$ used in 
several studies, defined as  
\beq
    \varepsilon_{mom} = \frac{m}{n_\n} 
    \frac{{\cal A}^{\n \p}}{{\cal B}^\n {\cal B}^\p - 
    \left({\cal A}^{\n \p}\right)^2} \ , \label{epsmom}
\eeq
is shown in Fig.~\ref{ent_param}. \ We set $c_\omega^2 = 7.148$ and 
$c_\sigma^2 = 12.684$ in what follows. \ These values are consistent with 
known nuclear matter properties and generate reasonable neutron star models 
\cite{glendenning97:_compact_stars,comer03:_rel_ent}.  

\section{Numerical Results} \label{results}

We consider three variations on the mode calculations: i) changing the master 
function, ii) including (or not) entrainment, and iii) varying the 
temperature.\ As we have already discussed, the main effect of the latter is 
to alter the size of the two-fluid layer. \ Since the parameter space is 
vast, and we are mainly interested in understanding the qualitative effects, 
we restrict our discussion to a few ``canonical'' background models. \ These  
determine the total mass $M$ and radius $R$ given a particular central baryon 
number density $n(0) = n_\n(0) + n_\p(0)$, obtained by first specifying 
$n_\n(0)$ and then finding $n_\p(0)$ through the condition of chemical 
equilibrium. \ Of course, as we vary the temperature $T$ of the star, the 
matter distribution is affected which, in principle, changes the values of 
$M$ and $R$. \ However, we find that these changes are negligible for the TOY 
model, and only slight for the PAL model (cf.~Table~\ref{ptable} and the 
disscusion in Sec.~\ref{gapsec}). \ 

\begin{figure}[t]
\centering
\includegraphics[width=8.5cm,clip]{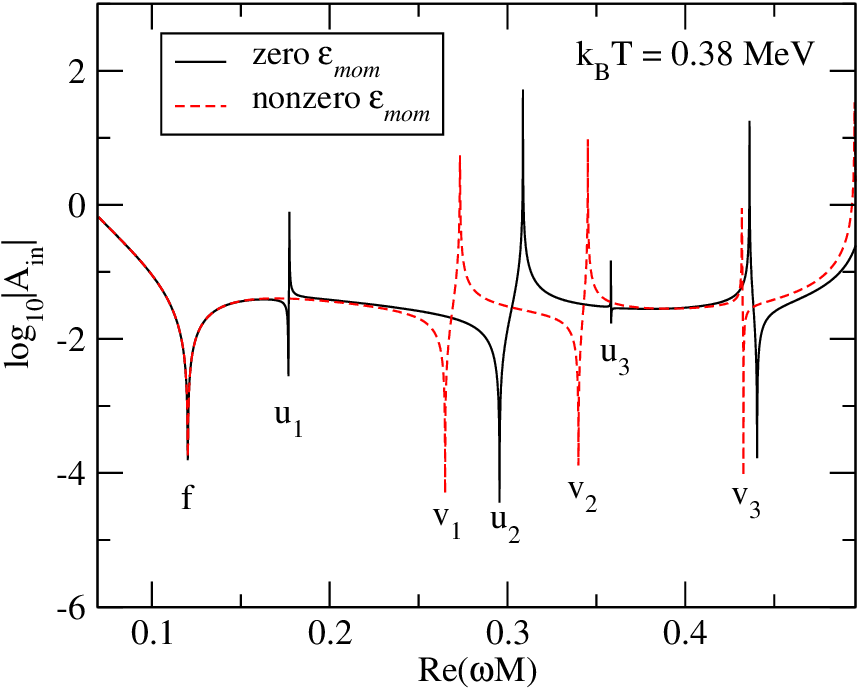}
\caption{Plot of ${\rm log}_{10}|A_{in}|$ ve ${\rm Re}(\omega M)$ for the 
TOY model, using $k_B T = 0.38$ MeV and $\varepsilon_{mom} = 0$ and 
$\varepsilon_{mom} \neq 0$.}
\label{Ain_comp2}
\end{figure}

\begin{figure}[t]
\centering
\includegraphics[width=8.5cm,clip]{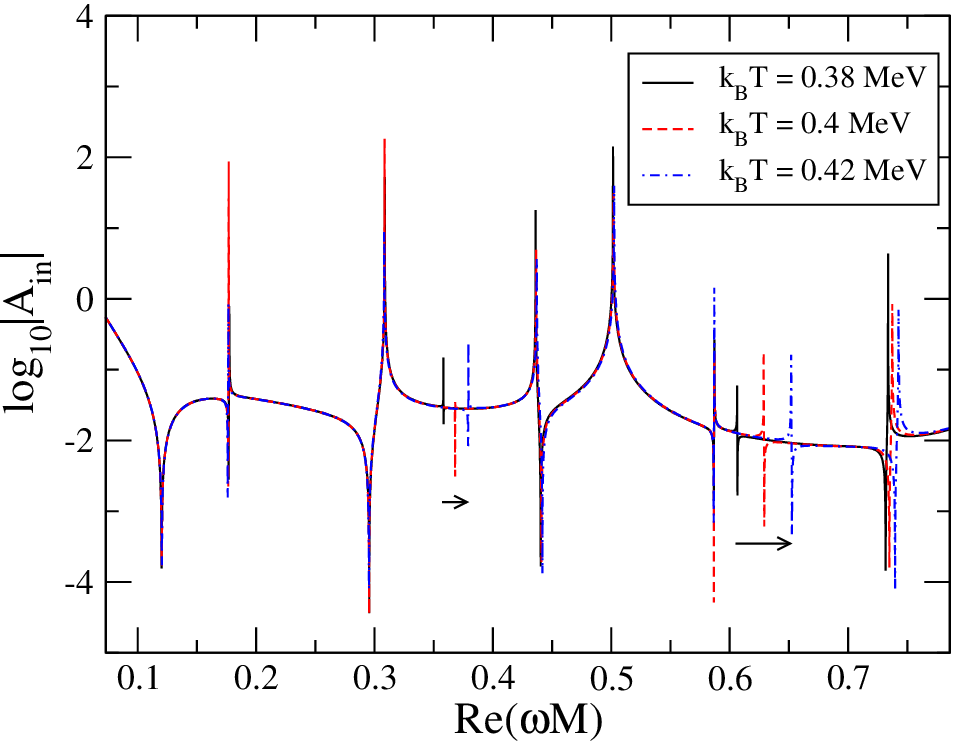}
\caption{Plot of ${\rm log}_{10}|A_{in}|$ vs ${\rm Re}(\omega M)$ for the TOY 
model, using $k_B T = \{0.38,0.4,0.42\}$ (MeV). \ Note the ``gap'' modes 
(indicated by arrows) that shift as the temperature (i.e.~the thickness of the 
two-fluid layer) is varied.}
\label{Ain_comp1}
\end{figure}

\begin{figure}[t]
\centering
\includegraphics[width=7.5cm,clip]{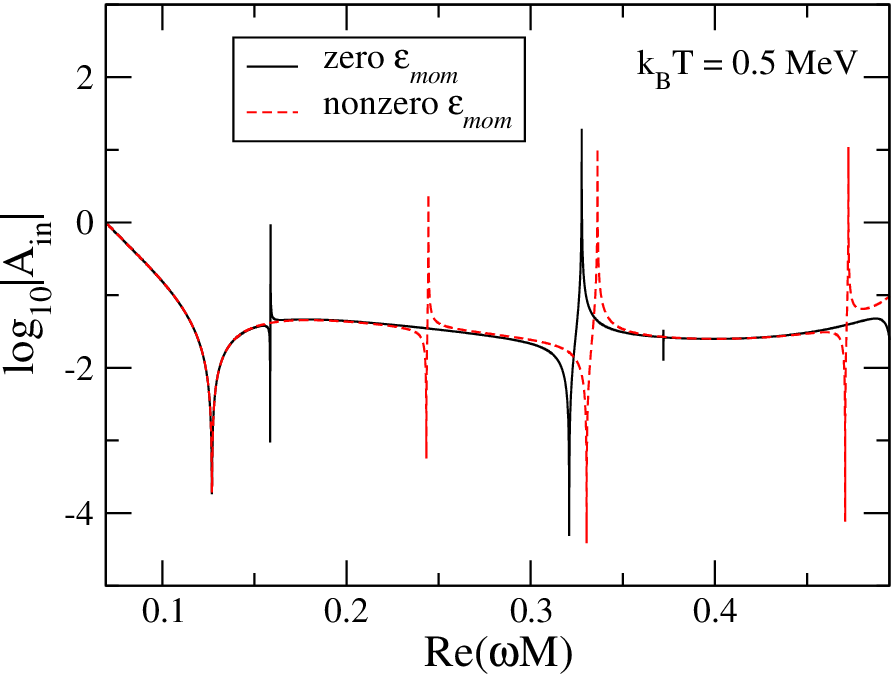}
\caption{Plot of ${\rm log}|A_{in}|$ vs ${\rm Re}(\omega M)$ for the PAL 
model, using $k_B T = 0.5$ MeV and for $\varepsilon_{mom} = 0$ and 
$\varepsilon_{mom} \neq 0$.}
\label{palent}
\end{figure}

\begin{figure}[t]
\centering
\includegraphics[width=7.5cm,clip]{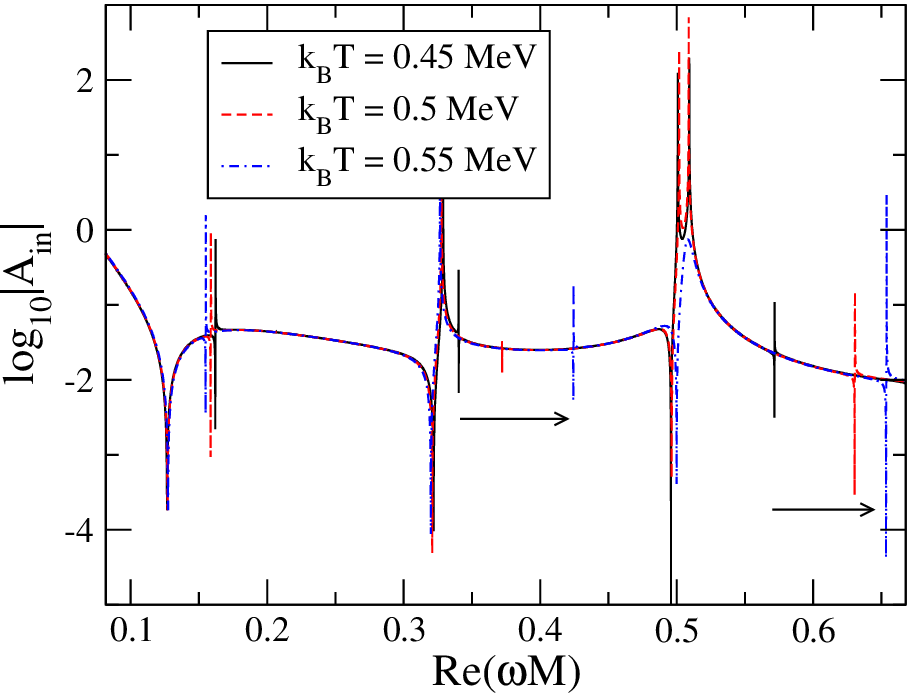}
\caption{Plot of ${\rm log}|A_{in}|$ vs ${\rm Re}(\omega M)$ for $k_B T = 
\{0.45, 0.5, 0.55\}$ (MeV) for the PAL model.}
\label{palt}
\end{figure}

\begin{figure}[t]
\centering
\includegraphics[width=7.5cm,clip]{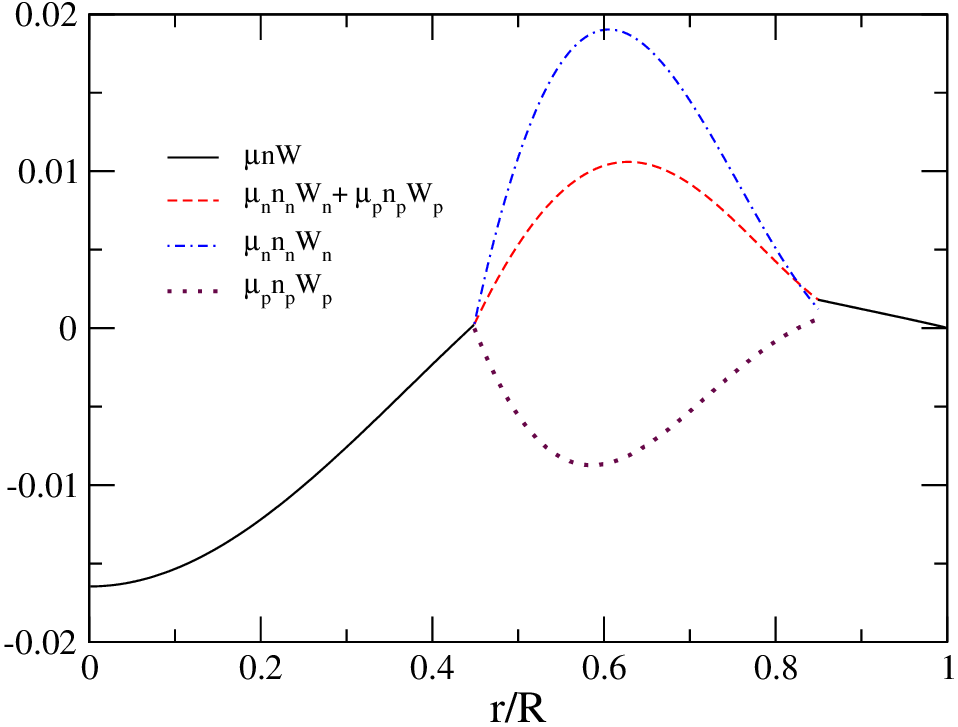}
\caption{Plot of the TOY model eigenfunction radial amplitude (vs $r/R$) for 
the first ``gap'' mode (i.e.~the arrow on the left in Fig.~\ref{Ain_comp1}) 
for the case $k_B T = 0.38$ MeV. \ The frequency of the mode is 
${\rm Re}(\omega M) = 0.3583$. \ The mode is such that the two fluids 
 exhibit counter-motion in the radial direction; i.e.~as the neutrons 
move in, say, the protons are moving out.}
\label{efunc2}
\end{figure}

\begin{figure}[t]
\centering
\includegraphics[width=7.5cm,clip]{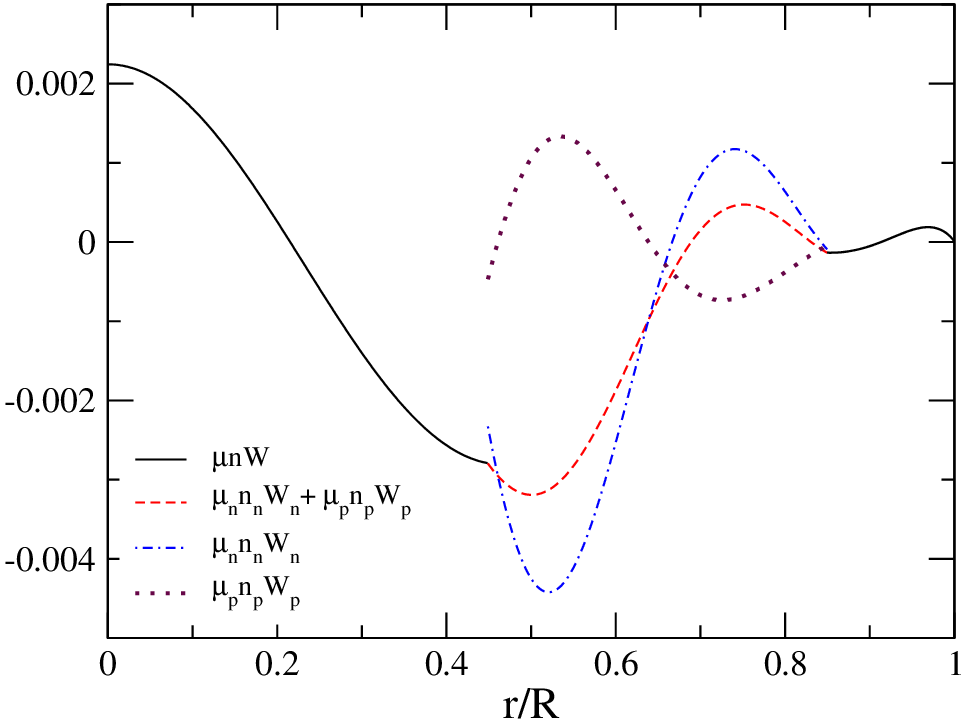}
\caption{Plot of the TOY model radial amplitude for the second ``gap'' mode 
(i.e.~the arrow on the right in Fig.~\ref{Ain_comp1}) for $k_B T = 0.38$ MeV. 
\ The frequency of the mode is ${\rm Re}(\omega M) = 0.6066$. \ The mode is 
such that the two fluids  exhibit counter-motion in the radial direction; 
i.e.~as the neutrons move in, say, the protons are moving out.}
\label{efunc3}
\end{figure}

\begin{figure}[t]
\centering
\includegraphics[width=10cm,clip]{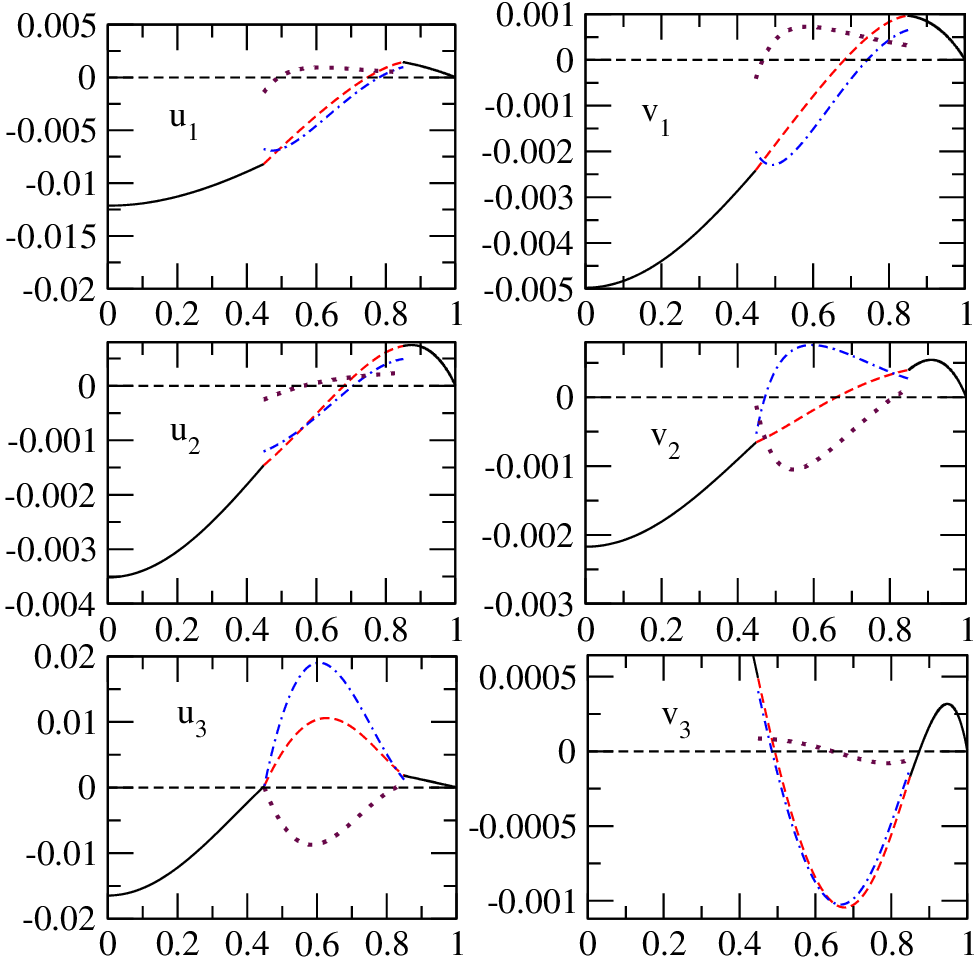}
\caption{Plot of the TOY model eigenfunction radial amplitudes (vs $r/R$) for 
the modes $u_1$..,$v_1$...etc labelled in Fig.~\ref{Ain_comp2}. \ The 
dashed, dot, and dot-dashed lines have the same denotations as given in the 
insets of the previous two figures. \ The remaining, horizontal line is to 
pinpoint where the amplitudes vanish.}
\label{efunc1}
\end{figure}

A quasinormal mode corresponds to a purely outgoing (gravitational) wave 
solution to the perturbation equations. \ The asymptotic  amplitude for 
ingoing waves, $A_{in}$, can therefore be used to locate the modes. \  
Figs.~\ref{Ain_comp2}, \ref{Ain_comp1}, \ref{palent}, and \ref{palt} show 
this amplitude versus frequency $\omega M$ for different models (see 
\cite{comer99:_quasimodes_sf} for the definition of $A_{in}$). \ The real 
part of the different quasinormal-mode frequencies can be approximated 
by the values that make $A_{in}$ vanish (or ${\rm log}_{10}|A_{in}|$ tend 
to $- \infty$). \ This technique does not provide the damping times 
of the modes due to gravitational-wave emission. \ In practice, we can use the 
technique described by Andersson et al 
\cite{andersson02:_oscil_GR_superfl_NS} to reliably determine the damping 
times. \ Our present discussion will, however, focus on the oscillation 
frequencies. \ Figs.~\ref{efunc2}, \ref{efunc3}, and \ref{efunc1} provide 
the radial profiles of mode eigenfunctions for the first few frequencies.

Several features of the mode spectrum are important to note. \ As in previous 
studies \cite{comer99:_quasimodes_sf,andersson02:_oscil_GR_superfl_NS} of the 
general relativistic two-fluid system, there are ``more modes'' than in the 
single-fluid case. \ This is expected since we have an additional fluid 
degree of freedom (albeit localised to a distinct layer in the star). \ From 
Fig.~\ref{Ain_comp2} we see that entrainment has a small effect on the first
mode in the spectrum, the f-mode, but clearly affects the next, a 
``superfluid'' mode, frequency by shifting it from the value labelled $u_1$ 
to that labelled $v_1$. \ Such behaviour is consonant with the local analysis 
of Andersson and Comer \cite{andersson01:_dyn_superfl_ns}, who show that the 
additional, superfluid modes depend on entrainment, while the fundamental 
f-mode does not. 

From Fig.~\ref{Ain_comp1} we see that as the size of the two-fluid layer is 
decreased most of the mode frequencies remain largely unchanged. \ However, 
in this model there are clear ``gap'' modes (emphasised by the arrows) whose 
defining characteristic is that the layer thickness (i.e.~the size of the gap 
energy $\Delta$ relative to the temperature $k_B T$) has significant impact 
on the frequency. \ As might be expected the frequencies increase as the 
thickness decreases. This is natural since the mode wavelengths must adjust 
to the size of the ``container''. 

For a semi-realistic determination of mode frequencies we consider 
Figs.~\ref{palent} and \ref{palt} which show ${\rm log}_{10}|A_{in}|$ vs 
$\omega M$ for the PAL master function. \ Qualitatively, Fig.~\ref{palent} 
shows that the effect of entrainment in the PAL model is the same as for 
the TOY model. \ Perhaps more interesting is the observation that the layer 
thickness affects {\em all} modes in the PAL model, cf.~Fig.~\ref{palt}. \ 
However, the most pronounced change is that the second ``gap'' mode 
(indicated by the arrow on the right of the figure) occurs earlier in the 
mode spectrum than in the case of the TOY model. \ Hence it seems clear that 
the combined background master function and the gap data are to some extent
distinguishable via the mode spectrum.

As expected, the  radial mode profiles have a richer structure than in the 
single-fluid case. \ There are also marked differences with models where the 
two fluids extend throughout the core
\cite{comer99:_quasimodes_sf,andersson02:_oscil_GR_superfl_NS}. \ In each of 
Figs.~\ref{efunc2}, \ref{efunc3}, and \ref{efunc1}, the solid line represents 
the total baryon flow in the one-fluid regions. \ In each two-fluid region 
the dot-dashed line represents the neutron flow, the dotted line represents 
the proton flow, and the dashed line represents the net baryon flow. \ Note 
that the boundary conditions between the one- and two-fluid regions imply 
that the net baryon flow must be continuous. 

In Figs.~\ref{efunc2} and \ref{efunc3} we see a characteristic feature of 
superfluid modes, the counter-motion between the neutrons and protons. \ The 
final Fig.~\ref{efunc1} illustrates the effect of entrainment. \ In the 
plot for the $v_3$ mode we see the beginnings of an ``avoided'' crossing. \ 
As first demonstrated by Andersson et al 
\cite{andersson02:_oscil_GR_superfl_NS}, when viewed as a function of the 
entrainment two neighbouring mode frequencies sometimes come close to crossing 
but veer off before intersection happens. \ Past this avoided crossing the 
two modes exchange character in the sense that a co-moving mode becomes 
counter-moving and vice versa.   

\section{Discussion: Next Generation Equations of State}
\label{neweos}

In this work we have extended the general relativistic two-fluid formalism 
used to model quasinormal-mode oscillations in such a way that superfluid 
neutron gap data can be incorporated. \ The main point is that the presence 
of a gap not only increases the number of fluid degrees of freedom locally; 
it also introduces a dynamical layering in the sense that the number of fluid 
degrees of freedom can change depending on whether the thermal energy of the 
star in a region is greater or less than the gap energy. \ It is clear that 
this structure will also be important for stellar rotational equilibria 
(eg.~as determined by a slow-rotation approximation).

In our models we analyzed the effects of different master functions, 
entrainment (obtained from a relativistic $\sigma-\omega$ mean field model), 
symmetry energy, and the presence of a gap. \ We found that the effect of 
entrainment on the mode spectrum is qualitatively consistent with earlier 
studies. \ On the other hand, the combined effects of the gap and the master 
function have distinguishable effects on the mode spectrum. \ Basically, our 
results illustrate that any ``realistic'' model for oscillating superfluid 
neutron stars must account not only for the ``bulk'' equation of state, one 
must also incorporate the entrainment and the superfluid gap data in a 
consistent way. \ We have, of course, not done this. \ At this point our 
models are more or less phenomenological. \ However, we have demonstrated 
that the computational technology required for this kind of study is now in 
place. \ What is missing is the input microphysics. 

At the present time there is, as far as we are aware, no single equation of 
state calculation that provides all the required parameters. \ This needs to 
change in the future. \ Furthermore, one must not forget the neutron star 
crust, where the lattice of nuclei will have elastic properties. \ These also 
need to be determined consistently. 

Let us discuss what (in our view) is required from the next generation 
equations of state if we want to model mode oscillations of multi-fluid 
neutron stars. \ To some extent these demands should be relatively 
straightforward to meet. \ Like most complex models, realistic equations of 
state are presented in tabular form. \ The question is what to include in the 
table. \ For traditional reasons, the information usually given is the 
pressure as a function of density and/or temperature. \ This would be adequate
if one were only interested in solving for a background configuration 
single-fluid star. \ The information is, however, not even complete for the 
oscillations of such models. \ In order to model the so-called g-modes, which 
are sensitive to both temperature and composition gradients, one would need 
also the various particle fractions. \ Moreover, the multi-fluid formalism 
presented here requires a number of additional quantities. 

In our experience, the two-fluid problem is most simply understood as having 
two independent thermodynamic parameters, the two number densitites 
$\{n_\n,n_\p\}$. \ We see from the background and perturbation equations in 
Section~\ref{review} that we require the set of values 
$\{\Lambda,\Psi,\B,\A,\partial \mu_\x/\partial n_\y,\Delta_\x\}$. \ Each  
variable is a function of $\{n_\n,n_\p\}$, and hence they will need to be 
given as two-dimensional arrays, the first index of each array standing for 
$n_\n$, say, and the second for $n_\p$. \ Of course, one usually assumes 
chemical equilibrium among the particle species on the background, which 
implies, say, that $n_\p = n_\p(n_\n)$. \ However, a number of the quantities 
that are needed for the perturbation analysis, like the entrainment $\A$, 
require information about the system away from equilibrium. \ If we let 
${\cal T}$ represent the set of thermodynamic variables, then a sample 
tabular equation of state could look something like that of Table 
\ref{tableeos}.

\begin{table}
 \begin{tabular}[t]{|c|c|c|c|c|c|}
 \hline 
            & $n_{\n,1}$ & $n_{\n,2}$ & $\dots$ & $n_{\n,N}$ \\
 \hline
 \ $n_{\p,1}$ & ${\cal T}_{1,1}$ & ${\cal T}_{2,1}$ & \dots & 
            ${\cal T}_{N,1}$ \\
 $n_{\p,2}$ & ${\cal T}_{1,2}$ & ${\cal T}_{2,2}$ & \dots & ${\cal T}_{N,2}$ \\
 \rotatebox{90}{$\dots$} & \rotatebox{90}{$\dots$} & \rotatebox{90}{$\dots$} 
    & \rotatebox{90}{$\dots$} & \rotatebox{90}{$\dots$} \\
 $n_{\p,N}$ & ${\cal T}_{1,N}$ & ${\cal T}_{2,N}$ & \dots & ${\cal T}_{N,N}$ \\
 \hline
 \end{tabular}
\caption{A schematic representation of a realistic, tabular equation of state 
that gives $N \times N$ entries for the set of thermodynamic variables 
${\cal T} = \{\Lambda,\Psi,\B,\A,\partial \mu_\x/\partial n_\y,\Delta_\x\}$.} 
\label{tableeos}
\end{table}

Can this information be provided within a single framework for determining
the ``equation of state''? \ We do not see why it should not be possible. \ 
Some of the quantities we require are already calculated, they are simply not 
presented in the final equation of state table. \ It should certainly be very 
easy to include the particle fraction in the tabulated data, and we see no 
reason why the entrainment coefficients should not be straightforward to 
determine as well. \ It is obviously the case that there is no universally 
agreed upon equation of state, or indeed method for its determination. \ From 
our point of view this is less relevant. \ We expect to live with 
uncertainties. \ After all, we do not have the luxury of having neutron stars 
readily available in the laboratory. \ The key point is that consistent 
models require all parameters to be consistent with a given microphysics 
calculation. \ Hopefully, an improved dialogue between neighbouring areas of 
research will allow us to make progress.

There are, of course, a number of related challenges. \ We have not yet 
satisfactorily answered the question of the minimal number of independent 
fluid degrees of freedom required to model a compact star. \  Glitch data 
tells us already that there are at least two. \ But what if we think about 
the deep core, quark deconfinement and CFL matter 
\cite{alford04:_review,alford00:_cfl}? \ Do the various Cooper pairings 
between quarks indicate the presence of independent ``fluids''? \ If they do, 
then what physical mechanisms exist to excite the additional degrees of 
freedom, and what is their physical interpretation? \ How can one hope to 
calculate the myriad of phases that could occur and then translate that into 
numerical models of rotating and oscillating multi-fluid compact stars? \ 
These are  challenging questions that require further consideration.

\acknowledgments
LML is supported in part by the Hong Kong Research Grants Council (grant 
numbers 401905 and 401807) and a postdoctoral fellow scheme at the Chinese 
University of Hong Kong. \ LML is also grateful for the hospitality of the 
Laboratoire de l'Univers et de ses Th\'{e}ories, Observatoire de Paris-Meudon 
where part of this work was done. \ NA gratefully acknowledges support from 
PPARC/STFC via grant numbers PP/E001025/1 and PP/C505791/1. \ GLC 
acknowledges partial support from NSF grant PHY-0457072. 

\section*{Appendix: Boundary conditions and numerical scheme}

We have three different regions inside the star: (1) one-fluid core; (2) 
two-fluid superfluid region; and (3) one-fluid envelope. \ In the one-fluid 
regions, $\tp$ is the baryon number density (with $\tchi$ being the chemical 
potential). \ In the superfluid region, $n_\n$ and $n_\p$ are the neutron and 
proton densities respectively ($\tp = n_\n + n_\p$). \ Note that, in this 
Appendix, we use tildes to make a distinction between the variables in the 
single-fluid and the multi-flud regime. \ $R_1$ and $R_2$ denote the 
core-superfluid and superfluid-crust interfaces, respectively. \ We let $R$ 
denote the star's radius. 

The one-fluid regions are governed by 4 ODEs for 
\beq
    \tilde{\bf{Y}} = \{ \tH1,\tK, \tWp, \tXp \} \ .
\eeq
The superfluid region is governed by 6 ODEs for 
\beq
    {\bf{Y}} = \{ H_1, K, W_\n, W_\p, X_\n, X_\p \} \ .
\eeq

\subsection{Boundary conditions}

At $r=0$, we have two conditions relating 
$\{ \tH1(0), \tK(0), \tWp(0), \tXp(0) \}$ (cf Eqs.~(A5)-(A6) of Comer et al 
\cite{comer99:_quasimodes_sf}, with $n_\n=0$, $W_\n(0)=0$, $n_\p 
\rightarrow \tp$, etc.~in our current notation). \ Explicitly, we have 
\beq
    \tXp(0) = \frac{e^{\nu_0/2}}{2} \overline{\tchi \tp} \tK(0) -
              \left(e^{\nu_0/2} \tilde{n}^2 
              \overline{\tilde{\rm B}^0_0 \tp} + 
              \frac{\omega^2}{l} e^{-\nu_0/2} \overline{\tilde{\cal B} 
              \tilde{n}^2}\right) \tWp(0) \label{eq:Xp0}
\eeq
and
\beq
    \tH1(0) = \frac{2}{l + 1} \tK(0) + \frac{16 \pi}{l (l +1 )} 
              \overline{\tchi \tp} \tWp(0) \ . \label{eq:H10}
\eeq

At each of the interfaces ($R_1$ and $R_2$), we have the following 4 
junction conditions according to (A16)-(A18) of Andersson et al 
\cite{andersson02:_oscil_GR_superfl_NS}:
\bea
    \tH1(R_c) &=& H_1(R_c) \ , \label{eq:H1} \\
    \tK(R_c) &=& K(R_c) \ , \\
    \tchi(R_c) \tp(R_c) \tWp(R_c) &=& \mu_\n(R_c) n_\n(R_c) W_\n(R_c) + 
    \mu_\p(R_c) n_\p(R_c) W_\p(R_c) \ , \\ 
    \tXp(R_c) &=& X_\n(R_c) + X_\p(R_c) \ , \label{eq:X}
\eea 
where we have assumed that $\Lambda$ and $\Psi$ are continuous across the 
interfaces. 

We further impose that the two fluids move in lock-step at the interfaces. \ 
This effectively translates to the condition 
\beq
    W_\n(R_c) = W_\p(R_c) \ . \label{eq:W}
\eeq
In summary, Eqs.~(\ref{eq:H1})-(\ref{eq:W}) are the required boundary 
conditions at the interfaces. 

Finally, at the surface of the star $r=R$, we have the single condition
\beq
    \tXp(R) = 0 \ .
\eeq

\subsection{Numerical scheme}

At $r=0$, we need only to specify $\{\tK(0),\tWp(0)\}$. \ The remaining 
variables $\{\tH1(0),\tXp(0)\}$ are determined by 
Eqs.~(\ref{eq:Xp0})-(\ref{eq:H10}). \ All of the second derivatives 
$\tH1''(0)$, $\tK''(0)$, etc.~are also determined. \ We choose two arbitrary 
values of $\{\tK(0),\tWp(0)\}$ and integrate the 4 ODEs from small $r_0$ up 
to the core-superfluid interface $r=R_1$. \ The general solution in the core 
region is 
\beq
     \tilde{\bf{Y}}(r) = \sum^2_{i=1} c_i \tilde{\bf{Y}}_i(r) \ , \ \
                         {\rm for}\ 0 \leq r \leq R_1 \ .
\eeq

Next we turn to the general solution in the superfluid region 
$R_1 \leq  r \leq R_2$. \ At $r = R_1$, the solution must satisfy the 
condition (\ref{eq:W}). \ This means that we must generate 5 linearly 
independent solutions. \ This is obtained by choosing five different sets of 
$\{H_1(R_1),K(R_1),W_n(R_1),W_p(R_1),X_n(R_1),X_p(R_1)\}$ (with 
$W_n(R_1) = W_p(R_1)$) and integrating to $r = R_2$. \ The general solution 
in this domain is 
\beq
    {\bf Y}(r) = \sum^7_{i=3} c_i {\bf Y}_i(r) \ , \ \
                 {\rm for}\ R_1 \leq r \leq R_2 \ .
\end{equation}
 
At the surface, our solution must satisfy $\tXp(R)=0$. \ Hence, we must 
generate 3 linearly independent solutions in the crust ($R_2 \leq r \leq R$). 
\ The general solution in this domain is 
\beq
\tilde{\bf Y} (r) = \sum^{10}_{i=8} c_i \tilde{\bf Y}_i (r) \ , \ \
                    {\rm for} \ R_2 \leq r \leq R \ .
\eeq 
 
After fixing the overall normalization by choosing the value of one of the 
$c_i$ (says, $c_{10}$), we have 9 remaining constants $c_i$ ($i = 1,..,9$)
to be determined by the boundary conditions at $R_1$ and $R_2$. 

First, at $r=R_1$, we have 4 conditions Eqs.~(\ref{eq:H1})-(\ref{eq:X}) to be 
satisfied. \ Note that the condition (\ref{eq:W}) has been used to generate 
the general solution in the superfluid region. \ Explicitly, these conditions 
become
\bea
    \sum^2_{i=1} c_i \tH1^{(i)} |_{R_1} &=& \sum^7_{i=3} c_i H_1^{(i)} |_{R_1} 
                 \ , \label{eq:match1} \\
    \sum^2_{i=1} c_i \tK^{(i)} |_{R_1} &=& \sum^7_{i=3} c_i K^{(i)} |_{R_1} 
                 \ , \\
    \sum^2_{i=1} c_i \left(\tchi \tilde{n} \tWp^{(i)}\right) |_{R_1} &=& 
                 \sum^7_{i=3} c_i \left(\mu_\n n_\n W_\n^{(i)} + \mu_\p n_\p 
                 W_\p^{(i)}\right) |_{R_1} \ , \\
    \sum^2_{i=1} c_i\tXp^{(i)} |_{R_1} &=& \sum^7_{i=3} c_i\left(X_\n^{(i)}
                 + X_\p^{(i)}\right) |_{R_1} \ ,
\eea
where we have defined ${\bf Y}_i = \{ H_1^{(i)},K^{(i)},W_n^{(i)},
W_p^{(i)},X_n^{(i)},X_p^{(i)} \}$ and 
$\tilde{\bf Y}_i = \{ \tH1^{(i)},\tK^{(i)},\tWp^{(i)},\tXp^{(i)} \}$.

Next, at $r=R_2$, we have 5 conditions Eqs.~(\ref{eq:H1})-(\ref{eq:W}):
\bea
    \sum^7_{i=3} c_i \left(W_\n^{(i)} - W_\p^{(i)}\right) |_{R_2} &=& 0 \ , \\
    \sum^7_{i=3} c_i H_1^{(i)} |_{R_2} &=& \sum^{10}_{i=8} c_i \tH1^{(i)} 
                 |_{R_2} \ , \\
    \sum^7_{i=3} c_i K^{(i)} |_{R_2} &=& \sum^{10}_{i=8} c_i \tK^{(i)} |_{R_2} 
                 \ , \\
    \sum^7_{i=3} c_i \left(\mu n W_\n^{(i)} + \mu_\p n_\p W_\p^{(i)}\right) 
                 |_{R_2} &=& \sum^{10}_{i=8} c_i \left(\tchi \tp \tWp^{(i)} 
                 \right) |_{R_2} \ , \\
    \sum^7_{i=3} c_i \left(X_\n^{(i)} + X_\p^{(i)}\right) |_{R_2} &=& 
                 \sum^{10}_{i=8} c_i \tXp^{(i)} |_{R_2} \ .\label{eq:match9}
\eea
The 9 equations Eqs.~(\ref{eq:match1})-(\ref{eq:match9}) can be used to 
determine the 9 constants $c_i$ ($i=1,..,9$). \ This completes the interior 
problem. 

\bibliography{biblio}

\end{document}